\documentclass[12pt,letterpaper]{article}
\usepackage[T1]{fontenc}
\usepackage[utf8]{inputenc}
\usepackage{lmodern}
\usepackage[margin=1in]{geometry}
\usepackage{amsmath,amssymb,amsthm,mathtools,mathrsfs,bm}
\usepackage{graphicx,xcolor,subcaption}
\usepackage{booktabs,float,multirow,tabularx}
\usepackage{enumerate,enumitem}
\usepackage[round]{natbib}
\usepackage{microtype}
\usepackage[hidelinks,unicode]{hyperref}
\hypersetup{
  pdftitle={Coherent and robust Bayesian inference under informative sampling via sample-level weight models},
  pdfauthor={Kosuke Morikawa; Jae Kwang Kim; Won Chang},
  pdfsubject={},
  pdfkeywords={Analytic inference; beta prime distribution; loss-likelihood bootstrap; Neyman orthogonality; Sverchkov--Pfeffermann identity}
}
\graphicspath{{figures/}}
\newtheorem{theorem}{Theorem}
\newtheorem{remark}{Remark}
\newtheorem{corollary}{Corollary}

\newtheorem{algo}{Algorithm}

\newtheorem{prp}{Proposition}

\numberwithin{equation}{section}

\newcommand{\bean}{\begin{eqnarray*}}
\newcommand{\eean}{\end{eqnarray*}}
\newcommand{\bea}{\begin{eqnarray}}
\newcommand{\eea}{\end{eqnarray}}

\title{\bfseries Coherent and robust Bayesian inference under informative sampling via sample-level weight models}
\author{%
  \normalsize Kosuke Morikawa\textsuperscript{1}\qquad
  Jae Kwang Kim\textsuperscript{1}\thanks{Correspondence: \texttt{jkim@iastate.edu}.}\qquad
  Won Chang\textsuperscript{2}\\[0.6em]
  \small\textsuperscript{1}Department of Statistics, Iowa State University, Ames, Iowa, USA\\
  \small\textsuperscript{2}Department of Statistics, Seoul National University, Seoul, Korea
}
\date{}

\begin{document}
\maketitle

\begin{abstract}

We develop two complementary Bayesian procedures for analytic inference under informative sampling. The first constructs a coherent likelihood by modeling the sample-level distribution of the design weight given the study variable and recovering the conditional inclusion probability through the identity of \citet{Sverchkov04}; it is the Bayesian counterpart of a recently proposed conditional-likelihood estimator, attains the model-based efficiency bound under the correctly specified sample-level joint model, and can be biased otherwise. The second uses the same weight model only as input to a Neyman-orthogonalized analog of the design-weighted score and applies the loss-likelihood bootstrap; the resulting credible intervals are asymptotically sandwich-correct under Poisson sampling, and the posterior remains centered at the truth under misspecification of the weight model, although the working model can still affect the first-order variance. A beta regression on the population-level inclusion probability, inducing a beta prime sample-level weight distribution, serves as the parametric default. A data-adaptive implementation of the second procedure estimates the efficient factor $\bar\pi(x,y)=1/E_p(W\mid x,y)$ by a design-weighted Gamma/log-link regression and uses it in a joint one-step loss-likelihood bootstrap for the analytic parameter; its validity requires no rate conditions on the nuisance fit. We illustrate both procedures with simulations and Canadian Workforce data.

\end{abstract}

\noindent%

\newpage

\section{Introduction}

Survey data are increasingly used not only for descriptive inference about totals and means but also for \emph{analytic inference}---inference about the parameters of a model assumed to govern the superpopulation rather than the particular finite population at hand \citep{Chambers2003,rao2015}. Analytic inference is more demanding because the sampling design is no longer ancillary: when the inclusion probability $\pi_i$ remains correlated with the study variable $y_i$ after conditioning on the available covariates, the design is \emph{informative} \citep{Sugden1984,Pfeffermann1998p,Pfeffermann1999}, and the sample distribution of $y$ differs systematically from its population counterpart.

Bayesian methods are an attractive vehicle for analytic inference because they accommodate prior information and deliver natural uncertainty quantification, but their treatment of informative sampling is less developed than in the descriptive case, because there is no obvious intrinsic likelihood. Three strategies have been proposed. \citet{savitsky2016} construct a pseudo-posterior by raising the weighted likelihood to a power---a simple and principled construction in the general Bayesian tradition of \citet{bissiri2016}, though the raw pseudo-posterior undercovers unless its spread is recalibrated \citep{williams2021}.
\citet{Wang2018} use the asymptotic normal distribution of the pseudo maximum-likelihood estimator as a working likelihood, inheriting its inefficiency. \citet{novelo2019} specify a joint model for the response and the weight at the population level and integrate out the unobserved population units; the approach is fully coherent but requires a population-level weight model that is rarely identified from sample data alone.

In this paper we develop two Bayesian procedures with complementary strengths. The first models the sample-level distribution of the design weight given the study variable and uses the identity of \citet{Sverchkov04} to recover the conditional inclusion probability from this model. The sample-level joint density of the study variable and the weight then combines with the analytic-model density into a coherent joint likelihood, and standard Markov chain Monte Carlo delivers the full joint posterior. For the weight model we build on the weight-smoothing framework of \citet{kim2013} and the conditional-likelihood estimator of \citet{kimwang2023}: a beta regression on the population-level inclusion probability induces a beta prime distribution on the sample-level weight, a natural parametric default. Procedure~I recasts this construction as a coherent Bayesian posterior and is in this sense the Bayesian counterpart of the estimator of \citet{kimwang2023}. It is consistent under correct specification of the weight model; under misspecification it can be biased, as the simulation study of Section~\ref{sec:simwork} illustrates.

The second procedure uses the same sample-level weight model but folds it into a Neyman-orthogonalized analog of the design-weighted score introduced by \citet{morikawa2025}. The orthogonalization makes the estimating equation first-order insensitive to estimation error in the weight-model parameter, and the estimator stays centered at $\theta_0$ under any admissible working weight model (Proposition~\ref{prp:nonlocal}); the working model still enters the sandwich variance, so the robustness concerns the posterior center, not the entire limiting posterior. We convert the estimating equation into a Bayesian procedure with the loss-likelihood bootstrap (LLB) of \citet{lyddon2019} in profile form: the weight-model nuisance parameter is held at its maximum-likelihood estimate and only the analytic-model score is resampled. The output is a posterior sample in the general Bayesian framework of \citet{bissiri2016} that carries the sandwich variance of the M-estimator automatically---no normal approximation, Markov chain Monte Carlo, or manual variance calculation.

The orthogonalized score is a family indexed by a smoothing factor $q(x,y)$.  The semiparametric efficient member uses $q=\bar\pi(x,y)=1/E_p(W\mid x,y)$, but this population-level conditional mean is not directly observed.  We therefore add a data-adaptive profile implementation: estimate $m_0(z)=E_p(W\mid Z=z)$, $Z=(X,Y)$, by a design-weighted Gamma/log-link regression, set $\widehat{\bar\pi}(z)=1/\widehat m_0(z)$, and apply the loss-likelihood bootstrap to the joint analytic parameter with this nuisance estimate held fixed.  This preserves the sample-based spirit of the paper while connecting Procedure~II to the efficient score of \citet{morikawa2025}.
Because the orthogonality identity is exact in the working factor rather than merely first-order,  no $o_p(n^{-1/4})$-type first-order nuisance-rate condition is imposed and a misspecified nuisance regression costs efficiency, not consistency or coverage (Proposition~\ref{prp:crossfit}).

The two procedures occupy distinct niches. Procedure~I delivers a coherent Bayesian likelihood and, under a correctly specified weight model, attains the model-based efficiency bound of the sample-level likelihood (Theorem~\ref{thm:procI-bvm}); it is the natural choice when the analyst is willing to commit to a parametric weight model. Procedure~II uses the weight model only as input to an orthogonalized estimating equation and is preferable when robustness to weight-model misspecification matters more than a coherent likelihood interpretation: it remains centered under any working weight model and delivers asymptotically calibrated sandwich credible intervals automatically, trading the contingent optimality of the parametric likelihood for that robustness.
To our knowledge, this is the first treatment of Bayesian inference under informative sampling to pair a coherent-likelihood posterior with a misspecification-robust orthogonalized-score posterior and to give an explicit account of their relative strengths.

The remainder of the paper is organized as follows. Section~2 establishes notation and reviews existing Bayesian approaches. Section~\ref{sec:procI} develops Procedure~I. Section~\ref{sec:llb} develops Procedure~II and its asymptotics. Section~\ref{sec:bnp-barpi} develops the data-adaptive implementation of the efficient factor $\bar\pi$ and establishes its validity.

Section~\ref{sec:simwork} reports a simulation study and Section~\ref{sec:LFS} illustrates both procedures on the Canadian Workforce data of \citet{fuller2009}. Section~\ref{sec:conclusion} concludes.

\section{Basic Setup}

Consider the problem of making inference about $\theta$ in the statistical model $f( y ; \theta)$. When an auxiliary covariate vector $x_i$ is observed for each unit, the analytic model is the conditional density $f( y \mid x; \theta)$; for clarity we suppress $x$ in the review below and reinstate it in the notation summary at the end of the section. Instead of obtaining a random sample, suppose that we have the following two-phase sampling structure.

\begin{enumerate}
    \item A finite population $\mathcal{F}_N = \{y_1, \cdots, y_N \}$ is a realization from the superpopulation model with density $f( y ; \theta)$.
    \item From $\mathcal{F}_N$ we obtain a probability sample and observe the realized values $y_i$ of the study variable in the sample.
\end{enumerate}

To formally define the setup, let $U=\{ 1, 2, \ldots, N\}$ be the index set of the finite population and let $I_i$ be the sampling indicator, with $I_i = 1$ if unit $i$ is selected and $I_i = 0$ otherwise. The probability $\pi_i=P \left( I_i =1 \mid i \in U \right)$ is called the \emph{first-order inclusion probability} and is known in probability sampling. If the study variable $y_i$ is not independent of $\pi_i$ conditional on the covariates used in the model, the design is informative \citep{Pfeffermann1999}, and analyses ignoring it can be biased.

The standard design-based approach is pseudo maximum likelihood: the estimator maximizes
    \begin{equation}
     l_w ( \theta) = \sum_{i \in A} w_i \log f( y_i  ; \theta) \label{1}
     \end{equation}
    over the parameter space $\Theta$, where $w_i=\pi_i^{-1}$ and $A$ is the index set of the sample, or equivalently solves the pseudo score equation
    \begin{equation}
    \hat{S}_w ( \theta) \equiv  N^{-1} \sum_{i \in A} w_i S( \theta; y_i ) = 0 \label{1b}
    \end{equation}
    where $S( \theta; y) = \partial \log f(y; \theta)/ \partial \theta$.  Asymptotic normality of the pseudo maximum likelihood estimator is established under regularity conditions by \citet{Binder1983} and \citet{rubin-bleuer2005}.

We now consider Bayesian inference in this setting. The main difficulty is that there is no intrinsic likelihood function under complex sampling. \citet{savitsky2016} proposed the pseudo posterior distribution
\begin{equation}
 p_w ( \theta \mid \mbox{data}) \propto \exp \{ c \cdot  l_w ( \theta)  \} \times p_0 ( \theta) ,
 \label{pos1}
 \end{equation}
where $l_w ( \theta)$ is defined in (\ref{1}), $c= (\sum_{i \in A} w_i)^{-1} n$, and $p_0 ( \theta)$ is a prior for $\theta$; the factor $\exp \{ c \cdot l_w ( \theta)\}$ serves as a pseudo likelihood. This is a general Bayesian posterior in the sense of \citet{bissiri2016}, with the weighted log-likelihood as its loss and the scalar $c$ as a global learning rate. Posterior consistency holds, but because a single scalar cannot reproduce the design-induced sandwich covariance the raw credible sets undercover; this is the sense in which it is not a proper likelihood for Bayesian inference \citep{monahan1992}. \citet{williams2021} repair the undercoverage by a post-hoc projection of the posterior draws that rescales and rotates them to the pseudo-MLE sandwich covariance. Procedure~II of this paper works in the same general Bayesian tradition but obtains the sandwich covariance by construction rather than by post-hoc adjustment, while Procedure~I retains a coherent generative likelihood.

 In Section~\ref{sec:procI} we develop a fully likelihood-based alternative that can be more efficient than pseudo-likelihood methods under correct specification of the parametric weight model; in Section~\ref{sec:llb} we develop a Bayesian-bootstrap alternative built on a Neyman-orthogonalized analog of the design-weighted score (\ref{1b}), whose credible sets attain asymptotically correct frequentist coverage automatically.

Throughout, $E_p$ denotes expectation under the population or superpopulation law before sampling, $E_s(\cdot)=E_p(\cdot\mid I=1)$ denotes sample-level expectation, and $E_{\mathcal D}$ denotes the joint design--superpopulation expectation, i.e. expectation over both $(X,Y,W)$ and the sampling indicator $I$; unsubscripted conditional expectations such as $E\{ q S \mid x\}$ are taken under the analytic model $f( y \mid x; \theta)$. Under Poisson sampling with $P(I=1\mid X,Y,W)=1/W$, the Horvitz--Thompson identity gives $E_{\mathcal D}\{I W h(X,Y,W)\}=E_p\{h(X,Y,W)\}$.

Table~\ref{tab:notation} in Appendix~\ref{app:weight-models} collects the notation used throughout the paper. {Three inclusion-probability objects must be kept distinct: the unit-level $\pi_i$, which may be degenerate or genuinely random given $( x_i, y_i)$, and the two conditional summaries
\[
\tilde \pi( x, y) = P( I = 1 \mid x, y) = \{ E_s( W \mid x, y)\}^{-1} , \qquad
\bar \pi( x, y) = \{ E_p( W \mid x, y)\}^{-1} .
\]
The first is \emph{sample-level}, recoverable from the observed sample through the Sverchkov--Pfeffermann identity; the second is its \emph{population-level} counterpart and the efficient factor of Section~\ref{sec:llb}.} Since $\tilde \pi = E_p( \pi \mid x, y)$ and $\bar \pi = 1/E_p( \pi^{-1} \mid x, y)$, Jensen's inequality gives $\bar \pi \le \tilde \pi$, with equality exactly when the inclusion probability is degenerate given $( x, y)$. Likewise the sample-level weight density $f_s( w \mid x, y; \eta)$, which is what the analyst models, differs from the population-level density $f_p( w \mid x, y; \eta)$ by an inclusion-probability tilt.

{\small

}

\section{Procedure I: conditional-likelihood posterior with a sample-level weight model}
\label{sec:procI}

We now develop the first of our two procedures. The construction starts from a coherent likelihood object---the sample-level joint density of the study variable and the design weight given covariates---and translates it into a full joint Bayesian posterior, not an empirical-Bayes or plug-in approximation: the weight model and the analytic model are coupled through a common normalizing integral, and posterior sampling proceeds by Markov chain Monte Carlo over the joint parameter space.

The construction has four steps: (i)~posit a parametric model $f_s( w \mid x, y; \eta)$ for the sample-level conditional distribution of the design weight; (ii)~recover the conditional inclusion probability $\tilde \pi( x, y; \eta) = 1/E_s( W \mid x, y; \eta)$ from this model through the Sverchkov--Pfeffermann identity; (iii)~substitute $\tilde \pi$ into the sample-level conditional density of $y$ given $( x, I = 1)$, which yields the conditional likelihood of $\theta$; and (iv)~combine the conditional likelihood with the weight-model likelihood to form a joint posterior for $( \theta, \eta)$. The remainder of the section carries out these steps.

As noted in Section~2, the main difficulty in implementing Bayesian inference under complex sampling is that there is no intrinsic likelihood function for the design-weighted data. To circumvent this, we consider the conditional likelihood function given by
\begin{equation}
L_C ( \theta) = \prod_{i \in A} f( y_i \mid I_i = 1; \theta) = \prod_{i \in A} \frac{\tilde{\pi} (y_i) f( y_i; \theta) }{ \int \tilde{\pi} (  y) f(y ; \theta) dy },
\label{3-1}
\end{equation}
where $\tilde{\pi} ( y) = P( I=1 \mid y) $
is the conditional inclusion probability. The sample likelihood function in (\ref{3-1}) has been considered in the (outcome-dependent) two-phase sampling literature with known $\tilde{\pi} (y)$ \citep[e.g.,][]{wang2009causal,tao2021two}. In the survey-sampling setting, weight smoothing for analytic inference originates with \citet{kim2013}; the conditional-likelihood construction (\ref{3-1}) built on a sample-level weight model is due to \citet{kimwang2023}; see also \citet{kim2013b}, Section~8.2. Procedure~I below recasts their conditional-likelihood estimator as a coherent Bayesian posterior.

In our problem the conditional inclusion probability $\tilde{\pi}(y)$ is unknown; only the unconditional inclusion probabilities $\pi_i$ are available in the sample. Rather than specifying a model for $E_p(\pi \mid y)$ directly, we model the sample-level distribution of the weight $w_i=\pi_i^{-1}$, given by $f( w \mid y, I=1)$. Modeling the sample-level distribution of the design weight, rather than the population-level distribution of the inclusion probability, was advocated by \citet{Pfeffermann1998p} and developed into an inferential framework by \citet{morikawa2025}; we adopt this framework here. By \citet{Sverchkov04},
\begin{equation}
 \tilde{\pi} (y) \equiv E_p( \pi \mid y) = \frac{1}{E_s( w \mid y; \eta) },  \label{eq:2-2}
 \end{equation}
 for some $\eta$, so $\tilde{\pi}(y)$ can be recovered from the sample-level weight model. The weight model can be parametric or nonparametric; we first consider a parametric model with parameter $\eta$.

 Once $\tilde{\pi} (y; \eta)$ is obtained from (\ref{eq:2-2}), the conditional likelihood function in (\ref{3-1}) becomes
 \begin{equation}
L_C ( \theta \mid \eta) =  \prod_{i \in A} \frac{\tilde{\pi} (y_i; \eta) f( y_i; \theta) }{ \int \tilde{\pi} (  y; \eta) f(y ; \theta) dy } .
\label{3-1b}
\end{equation}
Here, $\theta$ is the parameter of interest and $\eta$ is the nuisance parameter.
When the analytic model is conditional on covariates $x$ and the weight model conditions on $(x, y)$ rather than $y$ alone, the same construction applies with $\tilde \pi( y; \eta)$ replaced by $\tilde \pi( x_i, y; \eta)$ and the integral taken over $y$ given $x_i$:
\begin{equation}
L_C( \theta \mid \eta) = \prod_{i \in A} \frac{\tilde \pi( x_i, y_i; \eta) f( y_i \mid x_i; \theta)}{\int \tilde \pi( x_i, y; \eta) f( y \mid x_i; \theta) \, dy} .
\label{3-1c}
\end{equation}
The form (\ref{3-1b}) is recovered as the special case $\tilde \pi( x, y) = \tilde \pi( y)$.

The proposal is the joint posterior obtained from the sample-level joint density
\begin{equation}
f_s( y_i, w_i \mid x_i; \theta, \eta) \;=\; f_s( y_i \mid x_i; \theta, \eta) \cdot f_s( w_i \mid x_i, y_i; \eta) ,
\label{joint-likelihood}
\end{equation}
where the sample-level conditional density of $y$ given $x$ follows from Bayes' rule under the sampling indicator,
\begin{equation}
f_s( y \mid x; \theta, \eta) \;=\; \frac{\tilde \pi( x, y; \eta) \, f( y \mid x; \theta)}{\int \tilde \pi( x, y'; \eta) f( y' \mid x; \theta) \, dy'} ,
\label{f-s-conditional}
\end{equation}
and $f_s( w \mid x, y; \eta)$ is the sample-level weight model. A natural choice is the beta-prime density (\ref{5b}) below. The corresponding joint posterior is
\begin{equation}
p( \theta, \eta \mid \mbox{data}) \;\propto\; \prod_{i \in A} f_s( y_i, w_i \mid x_i; \theta, \eta) \cdot p_0( \theta, \eta) ,
\label{joint-posterior}
\end{equation}
which factorizes through (\ref{joint-likelihood}) into the conditional-likelihood factor in $\theta$ and the weight-model factor in $\eta$, coupled through the normalizing integral in (\ref{f-s-conditional}).

The full conditional distributions are
\begin{align}
p( \theta \mid \eta, \mbox{data}) &\;\propto\; \prod_{i \in A} \frac{f( y_i \mid x_i; \theta)}{Z_i( \theta, \eta)} \cdot p_0( \theta) , \label{post-theta} \\
p( \eta \mid \theta, \mbox{data}) &\;\propto\; \prod_{i \in A} \frac{\tilde \pi( x_i, y_i; \eta)}{Z_i( \theta, \eta)} \cdot f_s( w_i \mid x_i, y_i; \eta) \cdot p_0( \eta) , \label{post-beta}
\end{align}
where $Z_i( \theta, \eta) = \int \tilde \pi( x_i, y; \eta) f( y \mid x_i; \theta) \, dy$ is the per-observation normalizing integral, and where the $\tilde \pi( x_i, y_i; \eta)$ factor in the numerator of $p( \theta \mid \eta, \mbox{data})$ has been absorbed into the proportionality constant because it does not depend on $\theta$. Posterior sampling proceeds by Markov chain Monte Carlo on (\ref{post-theta})--(\ref{post-beta}); sampler details are given in Appendix~\ref{app:procI-sampler}; the Gamma/log-link application of Section~\ref{sec:LFS} admits a closed-form simplification.

The information channel that distinguishes (\ref{joint-posterior}) from a plug-in posterior is the $\theta$-dependence of $Z_i( \theta, \eta)$ in (\ref{post-beta}): the analytic-model parameter $\theta$ informs $\eta$ through the normalizing integral, and conversely the weight model informs $\theta$ through $\tilde \pi$. This bidirectional coupling is the defining feature of the full joint Bayesian construction. A plug-in approximation fits the weight model by maximum likelihood from the marginal weight-model likelihood $\prod_{i \in A} f_s( w_i \mid x_i, y_i; \eta)$ and then targets only (\ref{post-theta}) at $\hat \eta$; it is substantially cheaper and corresponds to a modular (cut) posterior in the sense of \citet{plummer2015} and \citet{jacob2017}. Unlike the orthogonalized score of Section~\ref{sec:llb}, however, the conditional-likelihood score for $\theta$ is not in general Neyman-orthogonal to $\eta$, so the plug-in posterior is first-order equivalent to the full joint posterior only when the $\theta$--$\eta$ cross-information vanishes. We use the full joint posterior in both the simulation study of Section~\ref{sec:simwork} and the application of Section~\ref{sec:LFS}. In the latter, the Gamma/log-link weight model permits an exact posterior factorization after a simple reparameterization, allowing exact sampling of the Gaussian analytic block while retaining posterior uncertainty in $\eta$.

The beta-prime sample-level weight model of \citet{kimwang2023} is the parametric default used throughout. A beta regression \citep{ferrari2004} on the population-level inclusion probability, with mean $\mu \equiv E_p( \pi \mid y) = \tilde \pi( y; \beta)$ and precision $\phi$, induces a beta prime distribution for $w - 1$ at the population level, and the inverse-probability tilt of Remark~\ref{rmk:fp-form} in Appendix~\ref{app:betaprime-derivation} carries it to the sample level: under sampling, $w - 1$ again follows a beta prime distribution, with density
\begin{equation}
f( w \mid y, I=1) = \frac{ 1}{ B( \phi - \mu \phi, \mu \phi +1) } (w-1)^{(1-\mu) \phi -1} w^{-\phi-1} I( w> 1) .
\label{5b}
\end{equation}
The parameter is $\eta = ( \beta, \phi)$, and the mean of (\ref{5b}) is $E_s( W \mid y) = 1/\mu$, so the Sverchkov--Pfeffermann identity (\ref{eq:2-2}) returns $\tilde \pi( y; \eta) = \mu$ exactly: the population-level regression mean is recovered without approximation from the sample-level model. The derivation, together with extensions of the weight model to categorical study variables and stratified designs, is given in Appendix~\ref{app:weight-models}.

Procedure~I is a coherent parametric Bayesian posterior for the sample-level model $f_s$, so under correct specification its large-sample behavior follows from standard parametric asymptotics: the joint posterior over $( \theta, \eta)$ satisfies a Bernstein--von Mises property and the marginal posterior of $\theta$ concentrates at $\theta_0$, attaining the model-based efficiency bound of the sample-level likelihood. We state this as Theorem~\ref{thm:procI-bvm} and prove it, together with a comparison to the semiparametric bound relevant to Procedure~II, in Appendix~\ref{app:proof-thm-procI-bvm}.

\section{Procedure II: A general Bayesian procedure via the loss-likelihood bootstrap}
\label{sec:llb}

Procedure~I treats the weight model as a likelihood object, so its validity rests on that model. We now construct a procedure that uses the weight model where it helps but is, by construction, robust to its misspecification.

\subsection{The orthogonalized score}
\label{sec:orth-score}

Procedure~II retains the tilted density that underlies Procedure~I but treats it as a \emph{loss} rather than a likelihood. We develop the construction for a general working smoothing factor $q( x, y; \eta)$, positive and free of $\theta$; the choice of $q$ is taken up at the end of the subsection, after Proposition~\ref{prp:optimal-q} identifies the efficient member of the family.

Let $S( \theta; x, y) = \partial \log f( y \mid x; \theta) / \partial \theta$ be the analytic-model score and form the tilted density
\begin{equation}
g_q( y \mid x; \theta) = \frac{ q( x, y; \eta) \, f( y \mid x; \theta)}{ \int q( x, y'; \eta) f( y' \mid x; \theta) \, dy'} ,
\label{eq:linear-tilt}
\end{equation}
which has exactly the form of the Procedure~I sample-level density (\ref{f-s-conditional}), with the generic factor $q$ in place of the conditional inclusion probability. Differentiating its logarithm gives $\partial_\theta \log g_q = S( \theta; x, y) - m_q( x; \theta, \eta)$, where the centering
\begin{equation}
m_q( x; \theta, \eta) = \frac{E\{ q( X, Y; \eta) \, S( \theta; X, Y) \mid X = x; \theta \}}{E\{ q( X, Y; \eta) \mid X = x; \theta \}}
\label{eq:m-correction}
\end{equation}
is the $q$-weighted conditional mean of the score under the analytic-model density $f( y \mid x; \theta)$, arising as $\partial_\theta \log \int q f$. Raising the tilted density to the power $q$ defines the \emph{power tilt}
\begin{equation}
f_q( y \mid x; \theta) = g_q( y \mid x; \theta)^{\, q( x, y; \eta)} .
\label{eq:power-tilt}
\end{equation}
We do not normalize $f_q$: the exponent is a pointwise power, so $f_q$ is not a density; Procedure~II uses it as a loss, with per-unit contribution $w\, q( x, y; \eta) \log g_q( y \mid x; \theta)$, and the general Bayesian interpretation is developed in Remark~\ref{rmk:unified}. Since $q$ does not depend on $\theta$, the orthogonalized score per unit is
\begin{equation}
\psi^{\perp} ( \theta, \eta; x, y) = \partial_\theta \log f_q = q( x, y; \eta) \bigl\{ S( \theta; x, y) - m_q( x; \theta, \eta) \bigr\} .
\label{eq:psi-orth}
\end{equation}
The design weight enters at the estimating-equation stage: each sampled unit contributes $\psi^{\perp}$ with its Horvitz--Thompson factor $w_i$, giving the estimating function $\Psi_n^{\perp}( \theta, \eta) = \sum_{i \in A} w_i\, \psi^{\perp}( \theta, \eta; x_i, y_i)$.

The centering by $m_q$ is exactly what orthogonalizes the score against the nuisance. Because $m_q$ is the $q$-weighted conditional mean of $S$, the tilted score is conditionally centered under the analytic model,
\begin{equation}
E\bigl\{ q( X, Y; \eta) \bigl[ S( \theta; X, Y) - m_q( X; \theta, \eta) \bigr] \bigm| X = x; \theta \bigr\} = 0 \qquad \text{for every working } q .
\label{eq:orth-identity}
\end{equation}
This identity, immediate from the definition of the centering and requiring no external efficient-score characterization, is the engine of the construction: holding for every $q$ and every $\eta$, and combined with the Horvitz--Thompson identity, it makes $\Psi_n^{\perp}$ design-unbiased at $\theta_0$ for \emph{every} working factor. Two results follow: the limiting distribution of the estimator, valid under any working factor however misspecified (Proposition~\ref{prp:nonlocal}); and---because design-unbiasedness holds for every $\eta$---insensitivity to how the weight-model parameter is estimated (Corollary~\ref{cor:orth}).

We take the two in turn.

\begin{prp}[Asymptotic distribution of the orthogonalized-score estimator]
\label{prp:nonlocal}
Let $q( x, y; \eta)$ be a working smoothing factor with the weight-model parameter $\eta$ held fixed, and let $\hat \theta_q$ solve the orthogonalized estimating equation
\[
\Psi_n^{\perp}( \theta; \eta) = \sum_{i \in A} w_i\, \psi^{\perp}( \theta, \eta; x_i, y_i) = \sum_{i \in A} w_i\, q( x_i, y_i; \eta) \bigl\{ S( \theta; x_i, y_i) - m_q( x_i; \theta, \eta) \bigr\} = 0 ,
\]
with $m_q$ the centering (\ref{eq:m-correction}). Suppose the design is Poisson sampling with $w_i = 1/\pi_i$ the true inverse inclusion probabilities; the analytic model $f( y \mid x; \theta_0)$ is correctly specified; $\theta_0$ is the unique root of the population equation $E_{\mathcal D}\{ I\, W\, \psi^{\perp}( \theta, \eta; X, Y)\} = 0$, where $E_{\mathcal D}$ denotes the joint design--superpopulation expectation; and the matrices
\begin{equation}
{A_q = E_s\bigl\{ W q\, ( S - m_q)( S - m_q)^\top\bigr\}, \qquad V_q = E_s\bigl\{ W^2 q^2\, ( S - m_q)( S - m_q)^\top\bigr\} ,}
\label{eq:bread-meat}
\end{equation}
evaluated at $\theta_0$, are positive definite. Then, by the identity (\ref{eq:orth-identity}), $\theta_0$ is a root of the population equation for \emph{every} $\eta$, so $\hat \theta_q$ is consistent for $\theta_0$ under any working smoothing factor, however badly misspecified; and
\begin{equation}
\sqrt n\, ( \hat \theta_q - \theta_0) \rightsquigarrow N( 0, \Sigma_q), \qquad \Sigma_q = A_q^{-1} V_q A_q^{-1} .
\label{eq:sandwich-clt}
\end{equation}
\end{prp}

The proof is in Appendix~\ref{app:proof-thm-llb}. Consistency is non-local in the working factor---it survives arbitrary misspecification of $q$, not only factors near the truth---because the identity (\ref{eq:orth-identity}) makes $\theta_0$ a population root for every $\eta$. {The sandwich (\ref{eq:sandwich-clt}) is the standard M-estimator limit for the weighted per-unit score $w \psi^{\perp}$ of the sampled units, with bread and meat the sample-level moments (\ref{eq:bread-meat}): the meat carries $W^2$ against the bread's single $W$, the design-induced inflation.}

\begingroup

\endgroup

Because $q$ carries the weight-model parameter $\eta$, one might expect estimation of $\eta$ to disturb the limit (\ref{eq:sandwich-clt}). It does not, and this is the additional orthogonality property.

\begin{corollary}[Neyman orthogonality; insensitivity to weight-model estimation]
\label{cor:orth}
Under the conditions of Proposition~\ref{prp:nonlocal}, the population equation $E_{\mathcal D}\{ I\, W\, \psi^{\perp}( \theta_0, \eta; X, Y)\} = 0$ holds for \emph{every} admissible $\eta$, so the map $\eta \mapsto E\{ \Psi_n^{\perp}( \theta_0; \eta)\}$ is identically zero and its derivative vanishes,
\begin{equation}
\frac{\partial}{\partial \eta} E\bigl\{ \Psi_n^{\perp}( \theta_0, \eta) \bigr\} \bigg|_{\eta = \eta^*} = 0 ,
\label{eq:neyman-orth}
\end{equation}
where $\eta^*$ is the probability limit of $\hat \eta$ under the working weight model. Consequently, replacing the fixed $\eta$ by an estimator $\hat \eta \to_p \eta^*$ leaves the conclusion of Proposition~\ref{prp:nonlocal} unchanged: $\hat \theta_q$ evaluated at $\hat \eta$ has the same limiting distribution $N( 0, \Sigma_{q( \cdot; \eta^*)})$ as at the fixed $\eta^*$, so first-order estimation error in the weight model does not propagate to $\hat \theta_q$.
\end{corollary}

The property is exact rather than local: (\ref{eq:neyman-orth}) holds because the population score is identically zero in $\eta$, not merely tangent to zero at $\eta^*$. The route also departs from established constructions of Neyman-orthogonal scores, which debias additively, estimating auxiliary nuisance quantities and projecting the score onto the orthocomplement of the nuisance tangent space \citep{neyman1959,chernozhukov2018double}; here the orthogonality arises multiplicatively, from the pointwise power transformation (\ref{eq:power-tilt}), with no auxiliary estimation and no projection. The exactness is what licenses holding $\hat\eta$ fixed across Dirichlet draws in the profile loss-likelihood bootstrap of Section~\ref{sec:profile-llb}.

\begin{prp}[Optimal smoothing factor]
\label{prp:optimal-q}
Consider the orthogonalized score (\ref{eq:psi-orth}) with working factor $q( x, y)$, suppressing the fixed $\eta$ from the notation, and let $\Sigma_q = A_q^{-1} V_q A_q^{-1}$ be its sandwich variance from Proposition~\ref{prp:nonlocal}, with $A_q, V_q$ the bread and meat (\ref{eq:bread-meat}). Then $\Sigma_q$ is minimized in the Loewner order by
\begin{equation}
q^\star( x, y) = \bar \pi( x, y) = 1 / E_p( W \mid x, y),
\label{eq:qstar}
\end{equation}
at which $A_{q^\star} = V_{q^\star} = \mathcal I^\star := {E_s\{ W \bar \pi\, ( S - m_{\bar \pi})( S - m_{\bar \pi})^\top\}}$, so $\Sigma_{q^\star} = ( \mathcal I^\star)^{-1}$. The optimal score $\psi_{\bar \pi}$ coincides with the efficient score of \citet{morikawa2025}, Corollary~C.2, and attains the semiparametric efficiency bound.
\end{prp}

The proof is in Appendix~\ref{app:proof-optimal-q}. It rests on a meat--bread asymmetry, made explicit in the population forms of Remark~\ref{rmk:scaling} in the appendices: the design inflates the meat by $E_p( W \mid x, y)$ relative to the bread, so the optimal multiplier is the generalized-least-squares choice $q^\star = 1/E_p( W \mid x, y)$.

The efficient factor $\bar\pi=1/E_p(W\mid x,y)$ is a population-level conditional expectation and is not supplied directly by the sample-level conditional inclusion-probability model.  Proposition~\ref{prp:nonlocal} implies that replacing $\bar\pi$ by any working factor preserves the posterior center; Proposition~\ref{prp:optimal-q} shows the variance is minimized at $\bar\pi$.  We therefore keep the sample-level factor $\tilde\pi=1/E_s(W\mid x,y)$ as the simple default implementation of Procedure~II, and in Section~\ref{sec:bnp-barpi} develop a data-adaptive implementation that estimates $E_p(W\mid X,Y)$ by a design-weighted Gamma/log-link regression and plugs $\widehat{\bar\pi}=1/\widehat E_p(W\mid X,Y)$ into the joint one-step LLB.

\begin{remark}[A unified view of the two procedures]
\label{rmk:unified}
The two procedures are two uses of the same tilt. As noted at (\ref{eq:power-tilt}), $f_q$ is not a density; renormalizing it would re-center its score off the analytic model and destroy the identity (\ref{eq:orth-identity}). Procedure~II is accordingly the general Bayesian posterior
\[
\prod_{i \in A} g_q( y_i \mid x_i; \theta)^{\, w_i q_i} \, p_0( \theta) ,
\]
a power posterior of the Procedure~I likelihood with per-unit learning rate $w_i q_i$ (equal to $w_i \tilde \pi_i$ at $q = \tilde \pi$), whose spread the loss-likelihood bootstrap of Section~\ref{sec:profile-llb} calibrates to the sandwich variance. At the efficient factor $q=\bar\pi$, the sensitivity and variability
matrices coincide, $A_q=V_q$, and the orthogonalized estimator attains
the semiparametric efficiency bound. This information identity concerns
the estimating equation; it does not turn the power tilt into a normalized
likelihood. Procedure~II therefore remains a loss-likelihood-bootstrap
procedure rather than a coherent likelihood posterior.
\end{remark}

\subsection{Profile loss-likelihood bootstrap}
\label{sec:profile-llb}

Write $\hat \eta$ for the maximum-likelihood estimator under the marginal weight-model likelihood $\prod_{i \in A} f_s( w_i \mid x_i, y_i;\eta)$, and $\psi_i^{\perp}( \theta; \hat \eta) = \psi^{\perp}( \theta, \hat \eta; x_i, y_i)$ for the per-unit orthogonalized score (\ref{eq:psi-orth}) at $\eta = \hat \eta$. The profile loss-likelihood bootstrap holds $\hat \eta$ fixed across Dirichlet draws and resamples only the $\theta$-score:

\begin{algo}[Profile LLB for the orthogonalized score]
\label{algo:llb}
Compute $\hat \eta$ once as the maximizer of $L( \eta) = \prod_{i \in A} f_s( w_i \mid x_i, y_i;\eta)$. For $j = 1, \ldots, B$:
\begin{enumerate}
\item Draw Bayesian-bootstrap weights $( g_{j1}, \ldots, g_{jn}) \sim n\, \mathrm{Dirichlet}( 1, \ldots, 1)$, so that each $g_{ji}$ has mean $1$.
\item Compute $\theta^{( j)}$ as the solution to
\[
\sum_{i \in A} g_{ji} \, w_i \, \psi_i^{\perp} ( \theta; \hat \eta) = 0 .
\]
\end{enumerate}
Return $\{ \theta^{( j)} \}_{j=1}^B$ as posterior samples.
\end{algo}

Each iteration of Algorithm~\ref{algo:llb} is a weighted M-estimation problem in $\theta$ with composite weights $g_{ji} w_i$ multiplying the per-unit orthogonalized scores. For linear regression and canonical-link generalized linear models, $m_q( x; \theta, \eta)$ depends on $\theta$ only through the mean function $\mu( x; \theta) = E\{ Y \mid X = x; \theta\}$, and the orthogonalized estimating equation reduces to a weighted linear regression with weights $g_{ji} w_i \tilde \pi( x_i, y_i; \hat \eta)$ once the conditional expectation in (\ref{eq:m-correction}) is evaluated at a $\sqrt n$-consistent starting value of $\theta$; in practice the design-based pseudo-MLE serves this role. The one-step linearization is asymptotically equivalent to the iterative solution and inherits the same conditional CLT, by the standard one-step argument \citep[][Section~5.7]{Vaart:98}. The algorithm is embarrassingly parallel in $j$, requires no Markov chain Monte Carlo, and reuses any code already written to evaluate the orthogonalized score at the maximum-likelihood estimate $\hat \eta$. The profile construction is justified by the orthogonality property (\ref{eq:neyman-orth}): the variability of $\hat \eta$ contributes to the posterior variance of $\theta$ only at second order, so holding $\hat \eta$ at its full-sample value produces the same asymptotic posterior as a joint LLB over $( \theta, \eta)$.

\subsection{Asymptotic validity}

We now state the central theorem, in the asymptotic setup of \cite{isaki1982}. Proposition~\ref{prp:nonlocal} gives the sampling distribution of the point estimator $\hat \theta_n$; it remains to show that the loss-likelihood bootstrap posterior of Algorithm~\ref{algo:llb} reproduces that distribution, so that its credible sets are calibrated. We retain the conditions of Proposition~\ref{prp:nonlocal} with the default factor $q = \tilde \pi( \cdot; \hat \eta)$, so that $\eta^*$ is the probability limit of $\hat \eta$ (which may differ from any true parameter when the weight model is misspecified) and $\theta_0$ is the unique root of the orthogonalized population equation of Proposition~\ref{prp:nonlocal} at $\eta = \eta^*$.

\begin{theorem}[Asymptotic validity of the orthogonalized LLB posterior]
\label{thm:llb}
Suppose the conditions of Proposition~\ref{prp:nonlocal} hold with $q = \tilde \pi( \cdot; \eta^*)$; that $\hat \eta$ is a $\sqrt n$-consistent estimator of $\eta^*$ under the working weight model, so that by Corollary~\ref{cor:orth} evaluating the score at $\hat \eta$ leaves the limit unchanged; and that $\psi^{\perp}( \theta, \eta^*)$ is continuously differentiable in $\theta$ on a neighborhood of $\theta_0$. Then the posterior samples $\theta^{(j)}$ of Algorithm~\ref{algo:llb} satisfy, for almost every realization of the data and conditionally on the data,
\begin{equation}
\sqrt{n}\, \bigl( \theta^{(j)} - \hat \theta_n \bigr) \mid \text{data} \rightsquigarrow N\bigl( 0, \, \Sigma_{\tilde \pi} \bigr) ,
\label{eq:llb-clt}
\end{equation}
where $\hat \theta_n$ is the M-estimator solving $\Psi_n^{\perp}( \theta; \hat \eta) = 0$ and $\Sigma_{\tilde \pi} = A_{\tilde \pi}^{-1} V_{\tilde \pi} A_{\tilde \pi}^{-1}$ is its sandwich variance (\ref{eq:sandwich-clt}). The loss-likelihood bootstrap thus reproduces the frequentist sandwich of Proposition~\ref{prp:nonlocal}, and the credible sets it generates have asymptotically correct frequentist coverage.
\end{theorem}

In words, the orthogonalized LLB posterior is asymptotically Gaussian, centered at the M-estimator $\hat \theta_n$ and carrying its sandwich variance. Since $\hat \theta_n$ is consistent for $\theta_0$ under any working weight model (Proposition~\ref{prp:nonlocal}), the posterior concentrates at the true value even when the weight model is misspecified. The analyst computes no variance formula; the sandwich is implicit in the spread of the LLB draws.

\begin{remark}[Extension to complex designs]
\label{rmk:complex-designs}
The Poisson sampling assumption of Proposition~\ref{prp:nonlocal} reflects what the unit-level Dirichlet bootstrap of Algorithm~\ref{algo:llb} can deliver: the conditional CLT of \citet{lo1987} treats sampled units as approximately independent draws from a sample distribution, which is consistent with Poisson sampling. For designs with non-negligible second-order inclusion correlations---fixed-size designs that are not close to Poisson, stratified designs with within-stratum dependence, multi-stage designs with intra-cluster correlation---the unit-level Dirichlet draws in step~1 of Algorithm~\ref{algo:llb} should be replaced by a design-respecting analog: PSU-level Dirichlet weights for multi-stage designs, within-stratum Dirichlet weights for stratified designs, or a two-step bootstrap in the style of \citet{beaumont2009}. The orthogonalized score (\ref{eq:psi-orth}) itself does not require modification; only the resampling step in Algorithm~\ref{algo:llb} does.
\end{remark}

\begingroup

\section{A data-adaptive efficient implementation of Procedure II}
\label{sec:bnp-barpi}

Proposition~\ref{prp:optimal-q} identifies the efficient factor as the population-level conditional expectation $q^\star( z) = \bar\pi( z) = \{ E_p( W \mid Z = z)\}^{-1}$, $Z = ( X, Y)$, whereas the default implementation of Procedure~II uses the sample-level $\tilde\pi = 1/E_s( W \mid Z)$.  This section develops a data-adaptive implementation that estimates $m_0( z) = E_p( W \mid Z = z)$ from the sampled units and sets $\widehat{\bar\pi} = 1/\widehat m_0$, without a fully specified population-level weight distribution. We call this implementation DML--Bayes, where DML denotes double/debiased machine learning.

Since only sampled units are observed, the risk must be design-weighted.  For any positive function $m(z)=\exp\{f(z)\}$, consider the design-weighted Gamma quasi-likelihood risk
\begin{equation}
R_n(f)=\sum_{i\in A} w_i\left[w_i\exp\{-f(z_i)\}+f(z_i)\right]+\lambda J(f),
\label{eq:gamma-loglink-risk}
\end{equation}
where $J(f)$ is a roughness penalty.  By the Horvitz--Thompson identity its population counterpart is $E_p[ W \exp\{ -f( Z)\} + f( Z)]$, minimized pointwise at $\exp\{ f_0( z)\} = E_p( W \mid Z = z)$: the Gamma/log-link loss targets exactly the conditional mean needed for the efficient factor.  The positive-link formulation is more stable than a raw weighted square loss for $W$, because the fitted mean is constrained positive and the loss is matched to right-skewed design weights.

In the simulations below we implement (\ref{eq:gamma-loglink-risk}) by a generalized additive model with a bivariate thin-plate spline,
\begin{equation}
W_i\mid Z_i=z_i \quad\hbox{has mean}\quad m_0(z_i)=\exp\{f(z_i)\},
\qquad f(z)=s(x,y),
\label{eq:gamma-gam-barpi}
\end{equation}
fit on the sampled units by the function \texttt{gam} in the R package \texttt{mgcv} \citep{wood2011}
with \texttt{family = Gamma(link = ``log'')} and design weights $w_i$.  The smoothing parameter is selected by restricted maximum likelihood; kernel ridge or other positive-link regressions can be substituted, provided the fit is evaluated out of sample.

We use cross-fitting only for the nuisance estimate: split the sampled units into $K$ folds $I_1, \ldots, I_K$ (default $K = 2$), fit (\ref{eq:gamma-gam-barpi}) on $A \setminus I_k$, and set $\widehat{\bar\pi}_i = \widehat m_{-k}( z_i)^{-1}$ for $i \in I_k$.  Inference is based on the pooled cross-fitted score, not on fold-specific posteriors.

For the linear Gaussian analytic model we work with the joint parameter $\alpha = ( \theta_0, \theta_1, \tau)^\top$, $\tau = \log \sigma^2$, so that the error variance carries calibrated uncertainty as well.  Writing $S_\alpha$ for the joint analytic score and $\widehat m_\alpha( x; \alpha)$ for the $\widehat{\bar\pi}$-weighted centering, the joint-score analog of (\ref{eq:m-correction}), the  {cross-fitted weighted estimating contribution is denoted by $\widehat\varphi_{\alpha i}$, rather than $\widehat\psi_{\alpha i}$, to distinguish it from the unweighted orthogonalized score:}
\begin{equation}
\widehat\varphi_{\alpha i}(\alpha)=
w_i\widehat{\bar\pi}_i
\{S_\alpha(\alpha;x_i,y_i)-\widehat m_\alpha(x_i;\alpha)\}.
\label{eq:joint-dml-score}
\end{equation}
Explicit formulas for $S_\alpha$ and the centering integral, together with the Gauss--Hermite quadrature used to evaluate it, are presented in Appendix~\ref{app:gaussian-impl}.

\begin{algo}[DML--Bayes profile one-step LLB with Gamma/log-link $\widehat{\bar\pi}$]
\label{algo:dml-gamma-barpi}
\begin{enumerate}
\item Construct the cross-fitted $\widehat{\bar\pi}_i$ by fitting the weighted Gamma/log-link regression (\ref{eq:gamma-gam-barpi}) on the training folds.
\item Solve the pooled estimating equation $\sum_{i\in A}\widehat\varphi_{\alpha i}(\alpha)=0$ to obtain $\widehat\alpha$.
\item Compute the per-unit weighted contributions $\widehat\varphi_{\alpha i}=\widehat\varphi_{\alpha i}(\widehat\alpha)$ and derivatives
\[
\widehat A_i=
\left.\frac{\partial \widehat\varphi_{\alpha i}(\alpha)}{\partial\alpha^\top}\right|_{\alpha=\widehat\alpha}.
\]
The derivative includes that of the centering term $\widehat m_\alpha(x;\alpha)$ with respect to both $\theta$ and $\log\sigma^2$.
\item For each posterior draw $j$, generate Bayesian-bootstrap weights $(g_{j1},\ldots,g_{jn})\sim n\,\mathrm{Dirichlet}(1,\ldots,1)$ and compute the one-step draw
\begin{equation}
\alpha^{(j)}=
\widehat\alpha-
\left\{\sum_{i\in A}g_{ji}\widehat A_i\right\}^{-1}
\sum_{i\in A}(g_{ji}-1)\widehat\varphi_{\alpha i}.
\label{eq:onestep-alpha-draw}
\end{equation}
\item Return the $\theta$ components of $\{\alpha^{(j)}\}$ as posterior samples; transform $\tau^{(j)}$ to $\sigma^{2(j)}=\exp(\tau^{(j)})$ if inference on the error variance is desired.
\end{enumerate}
\end{algo}

Algorithm~\ref{algo:dml-gamma-barpi} is a profile general Bayesian procedure.  The nuisance regression $\widehat{\bar\pi}$ is held fixed after cross-fitting; the Bayesian variation comes from the loss-likelihood bootstrap applied to the orthogonalized analytic score.  The one-step formula (\ref{eq:onestep-alpha-draw}) is asymptotically equivalent to solving the Dirichlet-weighted estimating equation directly, while avoiding a new quadrature and nonlinear solve for every bootstrap draw.  In practice we also compute the corresponding sandwich interval from the same $\widehat\varphi_{\alpha i}$ and $\widehat A_i$ as a diagnostic; the LLB and sandwich intervals agree closely in the simulations.

The results of Section~\ref{sec:llb} do not cover Algorithm~\ref{algo:dml-gamma-barpi}: Proposition~\ref{prp:nonlocal} treats a fixed working factor and Corollary~\ref{cor:orth} a $\sqrt n$-consistent parametric $\hat \eta$, whereas the cross-fitted $\widehat{\bar\pi}$ is a data-dependent function whose convergence rate may be slower than $\sqrt n$. The next result closes this gap, and shows that the exactness of the orthogonality identity (\ref{eq:orth-identity}) {eliminates the usual $o_p(n^{-1/4})$-type first-order nuisance-rate requirement, once boundedness and $L_2(P)$ convergence are imposed}.

\begin{prp}[Validity of the cross-fitted one-step LLB]
\label{prp:crossfit}
Write $\psi^{\perp}( \alpha, q; x, y) = q( x, y) \{ S_\alpha( \alpha; x, y) - m_q( x; \alpha)\}$ for the joint-score analog of (\ref{eq:psi-orth}), with $m_q( x; \alpha) = E\{ q\, S_\alpha \mid x; \alpha\} / E\{ q \mid x; \alpha\}$ the $q$-weighted centering, and let the folds $I_1, \ldots, I_K$, with $K$ fixed, partition $A$ at random, independently of the data. Suppose:
\begin{enumerate}[label=(\roman*)]
\item the design is Poisson sampling with $w_i = 1/\pi_i$ the true inverse inclusion probabilities, the $\pi_i$ bounded away from zero; the analytic model $f( y \mid x; \alpha)$ is correctly specified; and $\alpha_0$ is the unique root of the population equation $E_{\mathcal D}\{ I\, W\, \psi^{\perp}( \alpha, q^*; X, Y)\} = 0$;
\item each cross-fitted factor $\hat q_{-k}$ is constructed from the units in $A \setminus I_k$ only, and there are constants $0 < c \le C < \infty$ and a fixed factor $q^*$ with $c \le \hat q_{-k}, q^* \le C$ and $\| \hat q_{-k} - q^*\|_{L_2( P)} \to_p 0$ for each $k$, where $P$ is the population law of $( X, Y)$;
\item $S_\alpha( \alpha)$ is continuously differentiable in $\alpha$ on a neighborhood $\mathcal N$ of $\alpha_0$ with
\[
E_p\{ \sup_{\alpha \in \mathcal N} ( \| S_\alpha( \alpha)\|^2 + \| \partial_\alpha S_\alpha( \alpha)\|)\} < \infty
\]
and
\[
\sup_x E\{ \| S_\alpha( \alpha_0)\|^2 \mid X = x\} < \infty,
\]
and the bread and meat $A_{q^*}, V_{q^*}$ of (\ref{eq:bread-meat}), computed at $q^*$ with $S_\alpha$ in place of $S$, are positive definite.
\end{enumerate}
Then:
\begin{enumerate}[label=(\alph*)]
\item \emph{(Exact debiasing; no first-order nuisance-rate condition.)} For every $k$ and every realization of $\hat q_{-k}$,
\begin{equation}
E_{\mathcal D}\bigl\{ I\, W\, \psi^{\perp}( \alpha_0, \hat q_{-k}; X, Y) \bigm| \hat q_{-k}\bigr\} = 0 ,
\label{eq:crossfit-exact}
\end{equation}
so the pooled estimating function of Algorithm~\ref{algo:dml-gamma-barpi} is exactly design-unbiased at $\alpha_0$ conditionally on the nuisance fits, whatever their quality.
\item The pooled estimator $\hat \alpha$ of step~2 of Algorithm~\ref{algo:dml-gamma-barpi} satisfies $\sqrt n\, ( \hat \alpha - \alpha_0) \rightsquigarrow N( 0, \Sigma_{q^*})$ with $\Sigma_{q^*} = A_{q^*}^{-1} V_{q^*} A_{q^*}^{-1}$.
\item Conditionally on the data, the one-step draws (\ref{eq:onestep-alpha-draw}) satisfy $\sqrt n\, ( \alpha^{( j)} - \hat \alpha) \mid \mathrm{data} \rightsquigarrow N( 0, \Sigma_{q^*})$, so the LLB credible sets are asymptotically calibrated at the limit  $q^*$.
\item If moreover $q^* = \bar \pi$, then $A_{q^*} = V_{q^*} = \mathcal I^\star$ and $\Sigma_{q^*} = ( \mathcal I^\star)^{-1}$: the procedure attains the optimum of the orthogonalized-score family, by the argument of Proposition~\ref{prp:optimal-q} applied verbatim to the joint score.
\end{enumerate}
\end{prp}

The proof is in Appendix~\ref{app:proof-crossfit}. Proposition~\ref{prp:crossfit} makes precise the sense in which the exact multiplicative orthogonality of Section~\ref{sec:orth-score} is stronger than the first-order Neyman orthogonality underlying double/debiased machine learning \citep{chernozhukov2018double}. In that framework the population score is only tangent to zero in the nuisance, the plug-in bias is quadratic in the nuisance error, and rate conditions of the form $o( n^{-1/4})$ on the nuisance estimator are load-bearing. Here, by claim (a), the population equation vanishes identically in the working factor  so the plug-in bias is exactly zero for every realization of $\widehat{\bar\pi}$; cross-fitting is needed only to decouple the fitted factor from the units on which it is evaluated and no $o_p(n^{-1/4})$-type first-order nuisance-rate condition enters; the boundedness and $L_2(P)$ convergence assumptions remain part of the asymptotic statement. The quality of the nuisance fit governs efficiency alone: claim~(b) holds at whatever $L_2$ limit the fitted factor has, and the bound is attained exactly when that limit is $\bar \pi$. In particular, if the Gamma/log-link regression (\ref{eq:gamma-gam-barpi}) is inconsistent for $E_p( W \mid Z)$, the one-step LLB posterior remains centered at the truth with calibrated spread at the factor it actually estimates; the analyst risks efficiency, not validity.
\endgroup

\begingroup

\section{Simulation Study}
\label{sec:simwork}

This section reports a simulation study with three goals.  {First, it confirms that Procedure~I attains the model-based efficiency benchmark under correct specification and that the default Procedure~II can recover nearly the same efficiency when the sample-level factor $\tilde\pi$ and the efficient factor $\bar\pi$ are close.}  Second, it shows the failure of likelihood-based weight modeling under a nonlinear observed selection surface that cannot be captured by a logit-linear beta-prime working model.  Third, it evaluates the data-adaptive DML--Bayes implementation of Procedure~II from Section~\ref{sec:bnp-barpi}.

Throughout, the analytic model is
\begin{equation}
y_i=\theta_0+\theta_1x_i+\epsilon_i,\qquad \epsilon_i\mid x_i\sim N(0,\sigma^2),
\label{eq:sim-analytic}
\end{equation}
with true values $\theta_0=0$, $\theta_1=1$, and $\sigma^2=0.5$.  We use finite populations of size $N=10,000$, Poisson sampling, and $500$ Monte Carlo replicates.  In both scenarios the inclusion probabilities are generated as $\pi_i\mid Z_i\sim\mathrm{Beta}(\mu_i\phi,(1-\mu_i)\phi)$, with the mean surface $\mu_i$ and precision $\phi$ specified per scenario below.  In each population the intercept of the inclusion-probability surface is calibrated so that the expected sample size is approximately $1,000$; the realized means are $1001$ in Scenario~A and $1000$ in Scenario~B. For numerical stability, both $\mu_i$ and the realized $\pi_i$ are truncated to $[10^{-6},1-10^{-6}]$ before sampling.

The methods compared are complete-case OLS (CC), the design-based pseudo-MLE (DB), the Wang--Kim--Yang asymptotic-Bayesian posterior (WKY), Procedure~I, Procedure~II with the default $\tilde\pi$ smoothing factor, the design-only loss-likelihood bootstrap (HT-LLB), the DML--Bayes Gamma/log-link implementation of Procedure~II (Section~\ref{sec:bnp-barpi}) with $K=2$ folds, an oracle version using the true $\bar\pi$, and the joint-model posterior of \citet{novelo2019} (LN\&S).  The oracle version is included only as a diagnostic for the efficiency bound; it is not available in practice.

For Procedure~I in the simulation study, we sample the full joint posterior in (\ref{joint-posterior}); implementation details are given in Appendix~\ref{app:procI-sampler}.

\subsection{Scenario A: correctly specifiable weight model}
\label{sec:simA}

We generate $x_i\stackrel{\mathrm{iid}}{\sim}N(0,1/2)$ and $y_i$ from (\ref{eq:sim-analytic}).  The inclusion-probability mean is
\begin{equation}
\mathrm{logit}(\mu_i)=\alpha_A+0.75x_i+0.5y_i,
\qquad \phi=2500,
\label{eq:simA-design}
\end{equation}
where $\alpha_A$ is the calibrated intercept.  The working beta-prime weight model used by Procedure~I, Procedure~II, and LN\&S has the same logit-linear form in $(x,y)$, so this scenario favors the parametric modeling procedures.

\begin{table}[t]
\centering
\caption{Simulation results: bias, root mean squared error, empirical 95\% interval coverage, and average interval length for $\theta_1$ across $500$ Monte Carlo replicates. Scenario~A: correctly specifiable weight model; Scenario~B: nonlinear observed selection surface.}
\label{tab:sim-A}\label{tab:sim-B}
\small
\setlength{\tabcolsep}{4.5pt}
\begin{tabular}{lrrrrrrrr}
\hline
 & \multicolumn{4}{c}{Scenario A} & \multicolumn{4}{c}{Scenario B} \\
\cline{2-5}\cline{6-9}
Method & Bias & RMSE & Cov. & Len. & Bias & RMSE & Cov. & Len. \\
\hline
CC (naive) & $-0.043$ & $0.054$ & $75\%$ & $0.129$ & $\phantom{-}0.308$ & $0.308$ & $0\%$ & $0.081$ \\
Design-based & $-0.006$ & $0.061$ & $91\%$ & $0.216$ & $\phantom{-}0.006$ & $0.072$ & $88\%$ & $0.229$ \\
WKY & $-0.006$ & $0.061$ & $91\%$ & $0.216$ & $\phantom{-}0.006$ & $0.072$ & $88\%$ & $0.229$ \\
Procedure~I & $-0.003$ & $0.034$ & $94\%$ & $0.131$ & $\phantom{-}0.487$ & $0.488$ & $0\%$ & $0.103$ \\
Procedure~II & $-0.004$ & $0.034$ & $94\%$ & $0.128$ & $-0.048$ & $0.432$ & $70\%$ & $0.296$ \\
HT-LLB & $-0.008$ & $0.059$ & $91\%$ & $0.207$ & $\phantom{-}0.010$ & $0.069$ & $86\%$ & $0.215$ \\
DML--Bayes Gamma--$\bar\pi$ & $-0.003$ & $0.033$ & $95\%$ & $0.131$ & $-0.001$ & $0.024$ & $94\%$ & $0.095$ \\
Oracle $\bar\pi$ & $-0.004$ & $0.034$ & $95\%$ & $0.131$ & $-0.001$ & $0.022$ & $94\%$ & $0.089$ \\
LN\&S & $-0.003$ & $0.034$ & $93\%$ & $0.130$ & $\phantom{-}3.968$ & $4.909$ & $0\%$ & $0.048$ \\
\hline
\end{tabular}
\end{table}

Table~\ref{tab:sim-A} shows that the data-adaptive efficient-factor implementation loses essentially no efficiency when the parametric weight model is correct: Procedure~I, Procedure~II, LN\&S, the oracle $\bar\pi$ implementation, and DML--Bayes all have RMSE about $0.033$--$0.034$ with close-to-nominal coverage, and the close agreement between the DML--Bayes and oracle rows reflects the accuracy of the Gamma/log-link regression for $E_p(W\mid X,Y)$ in this scenario.  The design-only methods remain unbiased but have almost twice the RMSE.

\subsection{Scenario B: nonlinear observed selection surface}
\label{sec:simB}

Scenario~B is designed to favor a flexible estimator of $\bar\pi$ while keeping all variables driving selection observed.  We again generate $x_i\sim N(0,1/2)$ and $y_i$ from (\ref{eq:sim-analytic}), but the inclusion-probability mean is a nonlinear and interactive function of the observed pair $(x_i,y_i)$:
\begin{equation}
\mathrm{logit}(\mu_i)=\alpha_B+2h_i,
\qquad \phi=500,
\label{eq:simB-design}
\end{equation}
where $\alpha_B$ is the calibrated intercept, and
\begin{align*}
r_i={}&1.25\sin(2.6x_i)+1.00\cos(1.8y_i)
+0.90(x_i^2-\overline{x^2})+0.90(y_i^2-\overline{y^2})\\
&+1.10(x_iy_i-\overline{xy})+0.75\sin(2x_iy_i)
+0.45\,1\{y_i>Q_{0.65}(Y),\ x_i>Q_{0.40}(X)\},
\end{align*}
with $h_i=(r_i-\bar r)/s_r$.  The working parametric weight model used by Procedure~I, Procedure~II, and LN\&S remains logit-linear in $(x,y)$ and is therefore misspecified.  In contrast, the DML--Bayes Gamma--$\bar\pi$ implementation fits a flexible bivariate smooth for $E_p(W\mid X,Y)$.

\begin{figure}[t]
\centering
\includegraphics[width=0.78\textwidth]{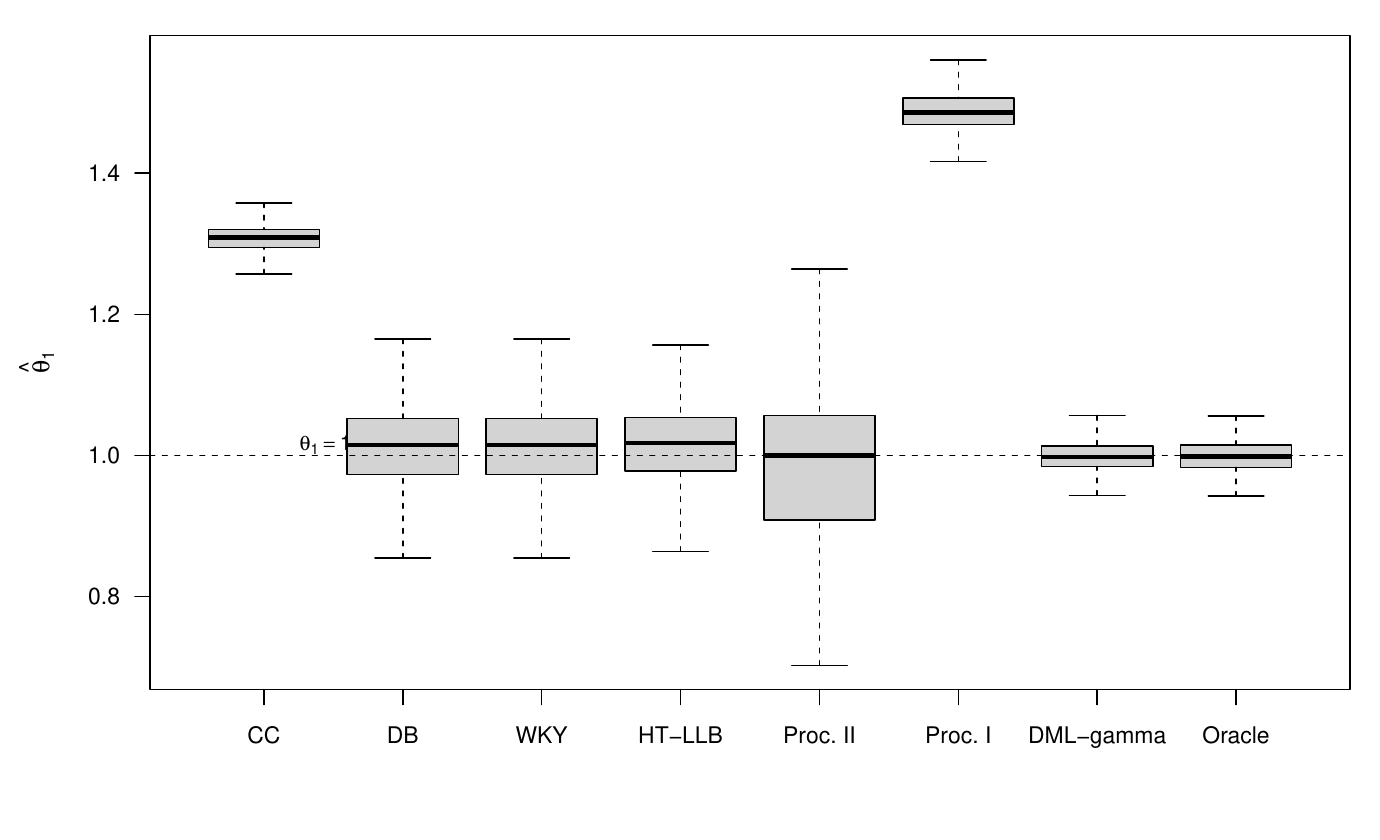}
\caption{Scenario~B: empirical distribution of $\hat\theta_1$ across $500$ Monte Carlo replicates.  The dashed horizontal line marks the true value $\theta_1=1$.  Outlying points are suppressed in the boxplots to emphasize the central distributions; the full RMSEs are reported in Table~\ref{tab:sim-B}.  LN\&S is omitted from the figure because its estimates are far outside the plotted range.}
\label{fig:sim-B}
\end{figure}

Table~\ref{tab:sim-B} and Figure~\ref{fig:sim-B} show that the likelihood-based weight-model methods are highly sensitive to this nonlinear misspecification: Procedure~I and LN\&S are badly biased.  The design-based methods remain centered but are relatively inefficient.  The default Procedure~II is centered for any fixed working factor (Proposition~\ref{prp:nonlocal}), but the price of a poor factor is variance (Proposition~\ref{prp:optimal-q}), and here the penalty is severe: the logit-linear factor is unstable in finite samples, and the resulting $70\%$ coverage shows that the asymptotic calibration of Theorem~\ref{thm:llb} has not set in at $n \approx 1000$.  The DML--Bayes Gamma--$\bar\pi$ implementation is both centered and efficient: its RMSE is $0.024$, close to the oracle $\bar\pi$ benchmark of $0.022$, and its coverage is near nominal.  The oracle comparison confirms that the efficient score itself delivers the expected gain; the small gap between oracle and DML--Bayes is due to estimating $E_p(W\mid X,Y)$ rather than knowing it.

Diagnostics support this interpretation.  In Scenario~A the correlation between $\widehat{\bar\pi}$ and the oracle $\bar\pi$ exceeds $0.998$ for both $K=2$ and $K=5$ folds.  In Scenario~B the corresponding correlation is about $0.992$, and the RMSE of $\widehat{\bar\pi}$ relative to the oracle is about $0.047$.  The effective sample size of the design weights in Scenario~B is only about $152$ despite a nominal sample size near $1000$, so the scenario is deliberately demanding.  The Gamma/log-link regression is nevertheless stable, whereas a raw weighted square-loss regression for $E_p(W\mid X,Y)$ was found to be much less stable in preliminary experiments.
\endgroup

LN\&S is included only as a benchmark. For the target slope, its chain failed at least one prespecified single-chain diagnostic criterion in 368 of 500 Monte Carlo replicates in Scenario~A and 499 of 500 replicates in Scenario~B; the latter count includes one initialization failure. All resulting finite estimates were retained in Table~\ref{tab:sim-A}.

\section{Application: Canadian Workforce data}
\label{sec:LFS}
\begingroup

We illustrate the procedures on the Canadian Workforce data analyzed by \citet{fuller2009}: $n = 142$ establishments selected by stratified random sampling from a frame of employers, with strata defined by employer size. Each record contains the design weight $w_i$, employment count, and total payroll; we take $y_i$ as log-payroll in dollars and $x_i$ as log-employment. The realized weights range from $12$ to $2826$, with quartiles $(32,39,743)$. The public file does not contain the stratum identifier, so the analysis uses only $(w_i,x_i,y_i)$: the stratified mixture model of Appendix~\ref{app:weight-models} is unavailable, and, following Remark~\ref{rmk:complex-designs}, the unit-level Dirichlet bootstrap treats the sampled establishments as approximately independent draws under the Poisson-sampling theory of Sections~\ref{sec:llb}--\ref{sec:bnp-barpi}.

The analytic model is the linear regression
\begin{equation}
y_i = \theta_0 + \theta_1 x_i + \varepsilon_i , \qquad \varepsilon_i \stackrel{\mathrm{iid}}{\sim} N( 0, \sigma^2) ,
\label{eq:LFS-analytic}
\end{equation}
relating log-payroll to log-employment at the establishment level. Complete-case OLS gives a slope of $0.907$, whereas the design-based pseudo-MLE gives $0.931$; together with the wide range of the weights, the gap indicates an informative design, modest in absolute terms but large enough that the treatment of the sampling design affects the substantive interpretation.

\subsection{Parametric sample-level weight model}

For Procedure~I and the default Procedure~II we specify the sample-level weight model as the Gamma GLM
\begin{equation}
W_i \mid x_i, y_i \sim \mathrm{Gamma}\bigl( \mu_i, \nu\bigr) , \qquad \log \mu_i = \beta_0 + \beta_1 x_i + \beta_2 y_i , \qquad \mu_i = E_s( W \mid x_i, y_i; \beta) .
\label{eq:LFS-wmodel}
\end{equation}
The fitted coefficients are $( \hat \beta_0, \hat \beta_1, \hat \beta_2) = (14.48,-0.52,-0.56)$ with dispersion $\hat\nu = 1.12$: the design weight is negatively associated with both employment and payroll, as expected under a size-stratified design. The rationale for preferring the Gamma GLM to the beta-prime default on these data is given in Appendix~\ref{app:LFS-wmodel}. For Procedure~I we sample the full posterior over $(\theta,\beta,\nu)$. Here $\tilde\pi(x,y;\beta)=\exp(-\beta_0-\beta_1x-\beta_2y)$, so the Gaussian normalizing integral is available in closed form, and a reparameterization factorizes the joint posterior into an exactly sampled Gaussian-regression block and a Gamma-weight-model block updated by adaptive random-walk Metropolis; details are given in Appendix~\ref{app:LFS-procI}. Procedure~II uses $\tilde\pi(x_i,y_i;\hat\beta)=1/\hat\mu_i$ in the orthogonalized score with the loss-likelihood bootstrap of Algorithm~\ref{algo:llb}; the design-only HT-LLB, using the bare Horvitz--Thompson score, serves as a benchmark.

\subsection{DML--Bayes Gamma--\texorpdfstring{$\bar\pi$}{bar-pi} implementation}

We also apply the data-adaptive efficient-factor implementation of Section~\ref{sec:bnp-barpi}: for each fold $k$, $m_0( z) = E_p( W \mid Z = z)$ is fitted on the complement of the fold by a bivariate smooth Gamma regression with log link and survey weights, and the fitted function is evaluated on the held-out fold to obtain $\widehat{\bar\pi}_{-k}( z_i)$. We then hold $\widehat{\bar\pi}$ fixed and apply the joint one-step LLB to $\alpha=(\theta_0,\theta_1,\log\sigma^2)$ as in Algorithm~\ref{algo:dml-gamma-barpi}. We use $K=2$ as the main analysis because the sample size is small, and $K=5$ as a sensitivity check.

\subsection{Results and interpretation}
\label{sec:LFS-comparison}

Table~\ref{tab:LFS-results} reports the point estimates, spread measures, and slope intervals. For the DML--Bayes rows, the intervals are the empirical $2.5\%$ and $97.5\%$ quantiles of the one-step LLB draws; the sandwich counterparts are nearly identical: $0.960$ with $(0.900,1.019)$ for $K=2$ and $0.952$ with $(0.889,1.016)$ for $K=5$.

\begin{table}[t]
\centering
\small
\caption{Canadian Workforce data: point estimates and spread measures. Entries in parentheses are posterior standard deviations for the Bayesian procedures and standard errors for complete-case OLS and the design-based pseudo-MLE. The interval column reports a $95\%$ interval for the slope $\theta_1$.}
\label{tab:LFS-results}
\begin{tabular}{lcccc}
\hline
Method & $\hat\theta_0$ (SD/SE) & $\hat\theta_1$ (SD/SE) & $95\%$ interval for $\theta_1$ & $\hat\sigma^2$ \\
\hline
Complete-case OLS & $10.019$ $(0.117)$ & $0.907$ $(0.032)$ & $(0.845,0.969)$ & $0.320$ \\
Design-based pseudo-MLE & $9.745$ $(0.128)$ & $0.931$ $(0.052)$ & $(0.829,1.033)$ & $0.299$ \\
Procedure~I & $9.839$ $(0.131)$ & $0.907$ $(0.032)$ & $(0.845,0.969)$ & $0.325$ \\
Procedure~II & $9.710$ $(0.166)$ & $0.948$ $(0.040)$ & $(0.868,1.024)$ & $0.258$ \\
HT-LLB & $9.745$ $(0.126)$ & $0.932$ $(0.050)$ & $(0.836,1.037)$ & $0.289$ \\
DML--Bayes Gamma--$\bar\pi$, $K=2$ & $9.701$ $(0.128)$ & $0.961$ $(0.032)$ & $(0.899,1.025)$ & $0.255$ \\
DML--Bayes Gamma--$\bar\pi$, $K=5$ & $9.723$ $(0.137)$ & $0.954$ $(0.034)$ & $(0.889,1.023)$ & $0.251$ \\
\hline
\end{tabular}
\end{table}

Relative to the complete-case slope of $0.907$, the design-based and design-only LLB slopes, $0.931$ and $0.932$, move the payroll--employment elasticity upward. Procedure~I gives essentially the same posterior mean slope as complete-case OLS, a structural consequence of the log-linear sample-level weight model: tilting a Gaussian conditional density by $\tilde\pi(x,y;\beta)\propto \exp(-\beta_2y)$ shifts the conditional mean by an $x$-free constant, so the marginal posterior for the slope is the standard Gaussian-regression posterior under the stated flat priors. Procedure~II and the DML--Bayes Gamma--$\bar\pi$ implementation both move the slope toward the design-adjusted side, with DML--Bayes giving the largest slope, $0.954$--$0.961$.

The main empirical message is that the DML--Bayes Gamma--$\bar\pi$ analysis combines design adjustment with smoothing. Relative to HT-LLB, the DML slope intervals are substantially shorter: HT-LLB gives $0.932$ with posterior SD $0.050$, whereas DML--Bayes gives posterior SDs $0.032$ and $0.034$ for $K=2$ and $K=5$. At the same time, two cautions apply. First, the design is stratified, the public file lacks the stratum identifier, and the theory of Sections~\ref{sec:llb}--\ref{sec:bnp-barpi} is Poisson-based, so the shorter DML intervals may reflect efficiency gain but may equally reflect unmodeled design structure. Second, $n = 142$ is small for the asymptotic calibration of Proposition~\ref{prp:crossfit} to be taken at face value. The application therefore suggests a potential efficiency gain from smoothing the design adjustment, with the numerical intervals interpreted as approximate; the DML intervals still overlap the design-based interval, and the agreement between $K = 2$ and $K = 5$ shows only that the conclusion is not driven by a particular sample split.

Figure~\ref{fig:LFS} shows the marginal LLB posterior densities of the slope for the two cross-fitting choices: the $K=2$ and $K=5$ densities are tightly concentrated, strongly overlapping, and centered to the right of the design-based estimate, consistent with the simulation evidence that estimating $\bar\pi=1/E_p(W\mid X,Y)$ by a positive-link regression can recover much of the efficiency gain without the rigidity of a parametric logit-linear weight model.

\begin{figure}[t]
\centering
\includegraphics[width=0.68\textwidth]{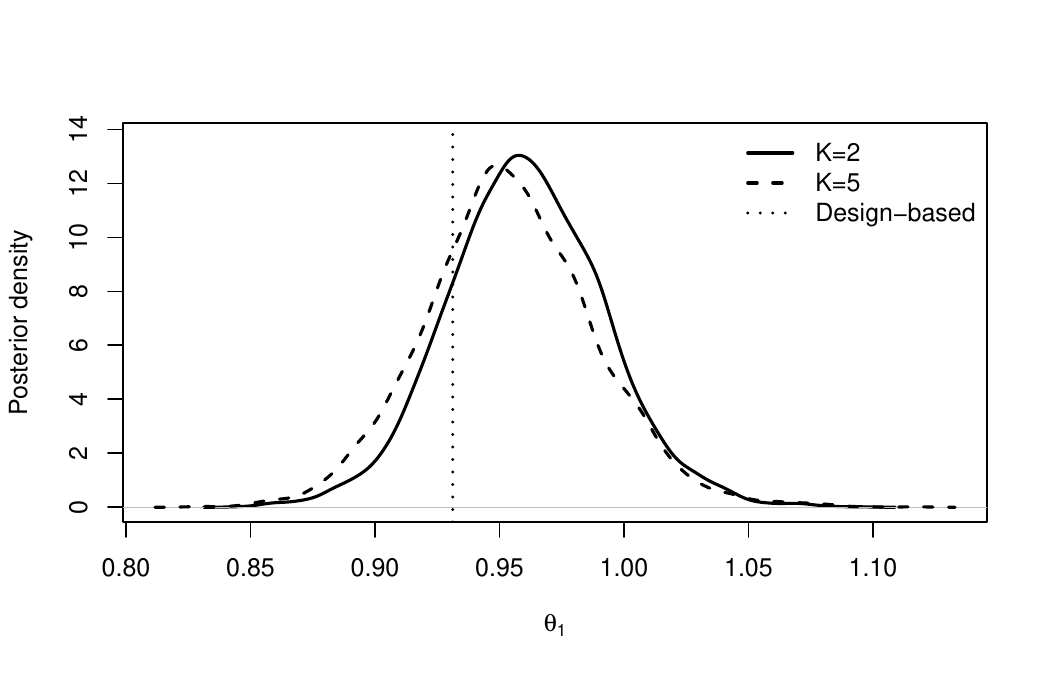}
\caption{Canadian Workforce data: marginal one-step LLB posterior densities of the slope $\theta_1$ for the DML--Bayes Gamma--$\bar\pi$ method with $K=2$ and $K=5$ cross-fitting folds. The vertical dotted line marks the design-based pseudo-MLE.}
\label{fig:LFS}
\end{figure}
\endgroup

\section{Concluding remarks}
\label{sec:conclusion}

We have developed two Bayesian procedures for analytic inference under informative sampling. Procedure~I delivers a coherent likelihood built from a parametric model for the sample-level weight distribution and, under correct specification, attains the model-based efficiency bound of that likelihood (Theorem~\ref{thm:procI-bvm}). Procedure~II folds the same weight model into a Neyman-orthogonalized estimating equation: estimation error in the weight-model parameter enters only at second order, and the posterior remains centered at $\theta_0$ under any admissible working weight model (Proposition~\ref{prp:nonlocal}), the working model affecting the sandwich variance but not the center.
The data-adaptive implementation completes the second procedure: estimating $\bar\pi=1/E_p(W\mid X,Y)$ by a design-weighted Gamma/log-link regression achieves near-oracle performance in the nonlinear simulation with asymptotically calibrated LLB uncertainty, requires boundedness and $L_2(P)$ convergence of the nuisance fit, but no $o_p(n^{-1/4})$-type first-order nuisance-rate condition (Proposition~\ref{prp:crossfit}), and in the Canadian Workforce application suggests a potential efficiency gain from smoothing the design adjustment.

The main limitations point to future work. The profile loss-likelihood bootstrap holds $\hat \eta$ fixed by an asymptotic orthogonality argument; a joint Dirichlet bootstrap over $( \theta, \eta)$ would propagate finite-sample nuisance uncertainty, at the cost of refitting the weight model within each draw. We have treated stratified but not multi-stage designs (Appendix~\ref{app:weight-models}); multilevel sample-level weight models would extend both procedures to intra-cluster correlation. Procedure~I uses flat priors for the analytic-model coordinates; in the simulation study, independent $N(0,10^2)$ priors on the weight-model coefficients and on $\log\phi$ regularize the full joint posterior under misspecification. Procedure~II uses the flat baseline prior implicit in the loss-likelihood bootstrap. A proper prior on $\theta$ is available through the calibrated general Bayesian posterior of \citet{bissiri2016} and \citet{lyddon2019}. The positive-link regression for $E_p(W\mid X,Y)$ in Section~\ref{sec:bnp-barpi} can be refined via adaptive kernels, additive-interaction decompositions, or multilevel smoothers for clustered designs. Further directions include modeling the sample-level weight distribution as a function of the analytic-model score, in the spirit of efficient score constructions, and Bayesian model averaging over weight models as an additional layer of robustness when the specification is uncertain.

\section*{Acknowledgments}
The research of Kim was partially supported by a grant from the U.S. National Science Foundation (2242820), a grant from the U.S. Department of Agriculture's National Resources Inventory, Cooperative Agreement NR203A750023C006, and Great Rivers CESU 68-3A75-18-504. The research of Morikawa was partially supported by JSPS KAKENHI Grant Numbers JP24K20743 and JP23H00466. The research of Chang was partially supported by the National Research Foundation of Korea (NRF) grant funded by the Ministry of Science and ICT (MSIT) (grant number RS-2025-00523567), the Global-LAMP Program of the National Research Foundation of Korea (NRF) grant funded by the Ministry of Education (grant number RS-2023-00301976), and the New Faculty Startup Fund from Seoul National University (grant number 326-20240027).

\clearpage
\appendix
\counterwithin{theorem}{section}
\counterwithin{remark}{section}
\counterwithin{table}{section}
\counterwithin{figure}{section}
\section*{Appendices}
\addcontentsline{toc}{section}{Appendices}
\noindent Appendix~\ref{app:weight-models} records the derivation of the beta-prime weight model and develops its categorical and stratified extensions; Appendix~\ref{app:proofs} contains proofs of the theoretical results; Appendix~\ref{app:LFS-calcs} collects implementation details and calculations for the numerical studies.

\section{The beta-prime weight model and extensions}
\label{app:weight-models}

\begin{table}[H]
\centering
\caption{Notation used throughout the paper.}
\label{tab:notation}
\small
\setlength{\tabcolsep}{4pt}
\begin{tabular}{llc}
\hline
Symbol & Meaning & Defined \\
\hline
$I_i$, $\pi_i$, $w_i$ & inclusion indicator, $\pi_i = P( I_i = 1)$, design weight $w_i = \pi_i^{-1}$ & Sec.~2 \\
$f( y \mid x; \theta)$, $S( \theta; x, y)$ & analytic (superpopulation) model and its score & Sec.~2 \\
$E_p( \cdot)$, $E_s( \cdot)$, $E_{\mathcal D}( \cdot)$ & population-model, sample-level $E_p( \cdot \mid I = 1)$, and joint & \\
& \quad design--superpopulation expectations & Sec.~2 \\
$\tilde \pi( x, y; \eta)$ & $P( I = 1 \mid x, y) = 1/E_s( W \mid x, y)$, sample-level & (3.2) \\
$\bar \pi( x, y)$ & $1/E_p( W \mid x, y)$, population-level & (4.9) \\
$f_s( w \mid x, y)$, $f_p( w \mid x, y)$ & sample- and population-level weight models  & (\ref{f-s-tilt}) \\
$g_q( y \mid x; \theta)$ & tilted analytic density with factor $q$ & (4.1) \\
$q( x, y; \eta)$, $m_q( x; \theta, \eta)$ & working smoothing factor; $q$-weighted conditional score mean & (4.2) \\
$\psi^{\perp}( \theta, \eta; x, y)$ & orthogonalized score $q\, \{ S - m_q\}$; $\Psi_n^{\perp} = \sum_{i \in A} w_i \psi_i^{\perp}$ & (4.4) \\
\hline
\end{tabular}
\end{table}

This appendix records the derivation of the beta-prime weight model stated in Section~3 and develops two extensions of the sample-level weight model---categorical study variables and stratified sampling. In each extension the working weight model is enlarged to accommodate the new structure, and either procedure can then be applied: Procedure~I substitutes the extended weight model into its conditional-likelihood posterior, while Procedure~II substitutes it into the orthogonalized score (4.4) and applies Algorithm~1.

\subsection{Derivation of the beta-prime model}
\label{app:betaprime-derivation}

Since $\pi$ is a probability, we consider the beta regression model \citep{ferrari2004}, written as
    \begin{equation}
     f( \pi \mid y)  \propto \pi^{\mu \phi -1} (1 - \pi)^{(1- \mu) \phi-1} ,
    \label{beta}
    \end{equation}
    where $ \mu \equiv E_p(\pi \mid y)= \tilde{\pi} (y; \beta)$
    and $\phi$ is the precision parameter satisfying
    $$ V( \pi \mid y) = \frac{ \mu( 1- \mu) }{ 1+ \phi } .$$
  From model (\ref{beta}) we obtain, for $w=1/\pi$,
\begin{equation}
f(w \mid y) \propto (w-1)^{(1-\mu) \phi -1} w^{-\phi}
\label{4}
\end{equation}
so $w-1$ follows a beta prime distribution with parameters $(1-\mu) \phi$ and $\mu \phi$.
Model (\ref{4}) is the population-level weight model, with parameter $\eta=(\beta, \phi)$.

To obtain the sample-level weight model, Bayes' formula gives
\begin{eqnarray}
 f( w \mid y, I=1) &\propto& P( I =1 \mid w, y) f( w \mid y) \notag
\\
&=& w^{-1} f( w \mid y) \notag \\
&\propto&    (w-1)^{(1-\mu) \phi -1} w^{-\phi-1} ,
\label{5a}
\end{eqnarray}
so in the sample $w-1$ again follows a beta prime distribution, with parameters $(1-\mu) \phi$ and $\mu \phi+1$.
Normalizing gives the sample-level density (3.10) stated in Section~3. Since the mean of a beta prime distribution with parameters $( a, b)$ is $a/( b - 1)$ for $b > 1$, the density (3.10) satisfies $E_s( W \mid y) = 1 + ( 1 - \mu) \phi / ( \mu \phi) = 1/\mu$, as used in Section~3.
\begin{remark}[An equivalent form via the population weight model]
\label{rmk:fp-form}
The joint sample-level density (3.5) can be re-expressed through the population weight density rather than the sample-level factorization (3.6). Write $f_p( w \mid x, y; \eta)$ for the population-level weight density and $f_s( w \mid x, y; \eta)$ for its sample-level counterpart appearing in (3.5); for the beta-prime default these are the model (\ref{4}) of Appendix~\ref{app:weight-models} and the density (3.10), respectively. The two are related by tilting with the inclusion probability,
\begin{equation}
 f_s( w \mid x, y; \eta) \;=\; \frac{ P( I=1 \mid x, y, w) \, f_p( w \mid x, y; \eta) }{ P( I=1 \mid x, y) } \;=\; \frac{ w^{-1} f_p( w \mid x, y; \eta) }{ E_p( W^{-1} \mid x, y; \eta) } ,
\label{f-s-tilt}
\end{equation}
using $P( I=1 \mid x, y, w) = 1/w$ pointwise. Since $E_p( W^{-1} \mid x, y; \eta) = E_p( \pi \mid x, y; \eta) = \tilde \pi( x, y; \eta)$ by the definition of $\tilde \pi$ in (3.2), the tilting relation (\ref{f-s-tilt}) is equivalent to the identity
\begin{equation}
 \tilde \pi( x, y; \eta) \, f_s( w \mid x, y; \eta) \;=\; w^{-1} f_p( w \mid x, y; \eta) .
\label{tilt-identity}
\end{equation}
Substituting (\ref{tilt-identity}) for the factor $\tilde \pi \, f_s$ in the numerator of the sample-level joint density (3.5)--(3.6) gives
\begin{eqnarray*}
 f_s( y, w \mid x; \theta,\eta)
   &=& \frac{ \tilde \pi( x, y; \eta) \, f( y \mid x; \theta) }{ \int \tilde \pi( x, y'; \eta) f( y' \mid x; \theta) \, dy' } \, f_s( w \mid x, y; \eta) \\
   &=& \frac{ f( y \mid x; \theta) \, w^{-1} f_p( w \mid x, y; \eta) }{ \int \tilde \pi( x, y'; \eta) f( y' \mid x; \theta) \, dy' } ,
\end{eqnarray*}
and hence the equivalent form of the joint posterior (3.7),
\begin{equation}
 p( \theta,\eta \mid \mbox{data}) = \prod_{i \in A} \left\{ \frac{ f( y_i \mid x_i; \theta) \, w_i^{-1} f_p( w_i \mid x_i, y_i;\eta) }{ \int \tilde \pi( x_i, y; \eta) f( y \mid x_i; \theta) \, dy } \right\} p_0( \theta,\eta) ,
 \label{3-6}
\end{equation}
which makes the multiplicative role of the design weight $w_i^{-1}$ explicit. The densities $f_p$ and $f_s$ differ by exactly the tilting factor in (\ref{f-s-tilt}), and it is this factor that supplies the $w_i^{-1}$ in (\ref{3-6}). The joint-model approach of \citet{novelo2019} works directly with the population weight density $f_p( w \mid x, y)$, which is more difficult to specify than the sample-level model $f_s( w \mid x, y; \eta)$ used here.
\end{remark}

\subsection{Application to categorical data}
\label{sec:categorical}

We now consider a categorical study variable $Y$ taking values in $\{ 1, \ldots, K\}$, with $\theta_k = P( Y = k)$ the population proportion of category $k$, marginal density
$$ p( y; \theta) = \prod_{k=1}^{K} \theta_k^{I( y=k)} , $$
and parameter space $\Theta = \{ ( \theta_1, \ldots, \theta_K); \sum_{k=1}^K \theta_k = 1, \theta_k > 0\}$. The per-category mean $\mu_k = E_p( \pi \mid y=k)$ is saturated---one free parameter for each of the $K$ categories---so the conditional inclusion probability $\tilde \pi( k) = 1/E_s( w \mid y=k)$ that enters the conditional likelihood is identified for any number of categories and cannot be misspecified in its mean. The beta-prime \emph{shape} within each category remains a parametric assumption, however: when $\mu_k$ is estimated jointly with the precision $\phi_k$ from the full beta-prime likelihood (\ref{5c}), within-category shape misspecification can still bias $\hat \mu_k$ relative to the saturated sample mean $1/\bar w_k$, where $\bar w_k$ is the average design weight among sampled units with $y=k$; estimating $\mu_k$ by $1/\bar w_k$ removes this sensitivity at the cost of not modeling the full within-category weight density. Misspecification can also enter through small-cell instability when some categories have few sampled units, and through the choice of categorization itself. The beta regression model in (\ref{beta}) is replaced by
  \begin{equation}
     f( \pi \mid y=k)  \propto \pi^{\mu_k \phi_k -1} (1 - \pi)^{(1- \mu_k) \phi_k-1} ,
    \label{beta2}
    \end{equation}
    where $ \mu_k = E_p(\pi \mid y=k)$
    and $\phi_k$ is the precision parameter satisfying
    $$ V( \pi \mid y=k) = \frac{ \mu_k ( 1- \mu_k) }{ 1+ \phi_k } .$$
The argument leading to (3.10) now gives
\begin{eqnarray}
 f( w \mid y=k, I=1)
&=&  \frac{ 1}{ B( \phi_k - \mu_k \phi_k, \mu_k \phi_k +1) }  (w-1)^{(1-\mu_k) \phi_k -1} w^{-\phi_k-1} ,
\label{5c}
\end{eqnarray}
and the posterior distribution over $( \theta, \mu, \phi)$, with $\mu = ( \mu_1, \ldots, \mu_K)$ and $\phi = ( \phi_1, \ldots, \phi_K)$, is
 \begin{eqnarray}
  p( \theta,  \mu, \phi \mid \mbox{data} ) & \propto &  \prod_{i \in A} \frac{1}{\sum_{k=1}^K \mu_k \theta_k} \prod_{k=1}^K \bigl\{ \mu_k \theta_k\, f( w_i \mid y_i = k, I = 1; \mu_k, \phi_k) \bigr\}^{I( y_i = k)} \nonumber \\
  & & \times\; p_0 ( \theta, \mu, \phi) .
  \label{4-5}
  \end{eqnarray}

\subsection{Stratified sampling}

Under stratified sampling it is natural to assume that $\pi$ follows a beta distribution within each stratum. Let $Z$ indicate the stratum to which the unit belongs. If $Z$ is observed in the sample, the weight model factorizes by stratum and the development of Appendix~\ref{app:betaprime-derivation} applies stratum by stratum; if $Z$ is unobserved, we use a mixture model.

The joint model for $( w, z, y)$ decomposes as
$$ [w, z, y] = [w | z, y ] \cdot [ z \mid y] \cdot [y] $$
where
\begin{eqnarray*}
 w^{-1} \mid (y, Z=h) &\sim &  \mathrm{Beta}( \mu_h \phi_h, (1-\mu_h)\phi_h) , \notag  \\
 P(Z=h \mid y) &= & \frac{ \exp (\alpha_{h0}+ \alpha_{h1}^\top y)}{ \sum_{h'=1}^H \exp ( \alpha_{h'0} + \alpha_{h'1}^\top  y)}
 \end{eqnarray*}
with $( \alpha_{10}, \alpha_{11}) = \mathbf{0}$ for identifiability.

  For Bayesian inference we use
      \begin{eqnarray}
  p( \theta, \alpha, \beta \mid \mbox{data} )
  \propto   \prod_{i \in A}  f( y_i, w_i  \mid   I_i=1; \theta, \alpha,  \beta )  p_0 ( \theta, \alpha, \beta) ,
  \label{post4}
  \end{eqnarray}
    where
  \begin{eqnarray*}
  f( y_i,  w_i  \mid   I_i=1; \theta,  \alpha, \beta )
  & =
 & f( y_i \mid  I_i=1; \theta, \alpha, \beta )\cdot  f( w_i \mid  y_i,  I_i=1; \alpha, \beta) .
  \end{eqnarray*}
   The first component is
    \begin{eqnarray*}
     f( y_i \mid  I_i=1; \theta, \alpha, \beta ) &=& \frac{ \tilde{\pi}( y_i; \alpha, \beta) f( y_i ; \theta) }{ \int \tilde{\pi}( y; \alpha, \beta ) f( y ; \theta) d y } ,
    \end{eqnarray*}
   where
   \begin{eqnarray*}
    \tilde{\pi}( y; \alpha, \beta) &=&  \sum_{h=1}^H \mu_h P( z=h \mid  y; \alpha) = \frac{ \sum_{h=1}^H \mu_h \exp (\alpha_{h0}+ \alpha_{h1}^\top  y)}{ \sum_{h'=1}^H \exp ( \alpha_{h'0} + \alpha_{h'1}^\top y)} .
    \end{eqnarray*}
  To compute the second component $f( w_i \mid  y_i,  I_i=1; \alpha, \beta)$, we use
   \begin{eqnarray}
    f( w_i \mid y_i,  I_i=1; \alpha, \beta)
   = \sum_{h=1}^H P( z=h \mid y_i, I_i=1) f( w_i \mid y_i, z=h, I_i=1) ,
   \label{18}
   \end{eqnarray}
   where
    \begin{eqnarray}
   P( z=h \mid y_i, I_i=1; \alpha, \beta)
  &=& \frac{ P(z=h \mid  y_i ; \alpha) \mu_h }{ \sum_{h'=1}^H P(z=h' \mid  y_i ; \alpha) \mu_{h'} }
  \label{multinom}
    \end{eqnarray}
    and $W-1\mid (y,z=h, I=1) \sim \mathrm{Beta'}((1-\mu_h)\phi_h, \mu_h\phi_h+1).$
Thus the sample-level weight model (\ref{18}) is a mixture of beta prime distributions with mixture probabilities (\ref{multinom}).

\section{Proofs}
\label{app:proofs}

\subsection{Proof of Proposition~1 and Theorem~1}
\label{app:proof-thm-llb}

Write $\Psi_n^{\perp}( \theta, \eta) = \sum_{i \in A} w_i\, \psi^{\perp}( \theta, \eta; x_i, y_i)$ with $\psi^{\perp}( \theta, \eta; x, y) = q( x, y; \eta)\{ S( \theta; x, y) - m_q( x; \theta, \eta)\}$ the per-unit orthogonalized score (4.4), and let $\hat \theta_q$ solve $\Psi_n^{\perp}( \theta; \eta) = 0$ at a fixed $\eta$; for Theorem~1 set $q = \tilde \pi$ and $\eta = \hat \eta$ with probability limit $\eta^*$, writing $\hat \theta_n$ for the resulting estimator.

\emph{Step~1: consistency and the sandwich CLT (Proposition~1).} Applying the Horvitz--Thompson identity to the conditional expectations in $m_q$ gives $E_{\mathcal D}\{ I W q S \mid X\} = E_p\{ q S \mid X\}$ and $E_{\mathcal D}\{ I W q\, m_q \mid X\} = m_q\, E_p\{ q \mid X\}$; the ratio definition of $m_q$---equivalently the identity (4.5)---cancels them, so
\[
E_{\mathcal D}\bigl\{ I\, W\, \psi^{\perp}( \theta_0, \eta; X, Y)\bigr\} = 0 \qquad \text{for every admissible } \eta .
\]
Thus $\theta_0$ is a root of the orthogonalized population equation under any working factor; taking it to be the unique root gives $\hat \theta_q \to_p \theta_0$ by the standard M-estimation argument \citep[][Theorem~5.7]{Vaart:98}, non-locally in $\eta$. {Writing $\mathbb P_n \{ w \psi^{\perp}\}( \theta, \eta) = n^{-1} \Psi_n^{\perp}( \theta, \eta)$ for the empirical mean over the sampled units, which, as in Step~1 of Appendix~\ref{app:proof-thm-procI-bvm}, are independent draws from the sample distribution, with $E_s\{ W \psi^{\perp}( \theta_0, \eta)\} = E_{\mathcal D}\{ I W \psi^{\perp}( \theta_0, \eta)\} / P( I = 1) = 0$ for every admissible $\eta$ by the displayed identity, and expanding $0 = \mathbb P_n \{ w \psi^{\perp}\}( \hat \theta_q, \eta)$ about $\theta_0$,}
\[
\sqrt n\, ( \hat \theta_q - \theta_0) = -A_q^{-1} \sqrt n\, \mathbb P_n \{ w \psi^{\perp}\}( \theta_0, \eta) + o_p( 1) \;\rightsquigarrow\; N( 0, \Sigma_q), \qquad \Sigma_q = A_q^{-1} V_q A_q^{-1} ,
\]
{where $A_q = -\partial_\theta E_s\{ W \psi^{\perp}( \theta_0, \eta)\}$ is the bread and $V_q = E_s\{ W^2 \psi^{\perp}( \theta_0, \eta)^{\otimes 2}\}$ the meat of (4.6), both positive definite by assumption; the meat is the second moment of the mean-zero per-unit summand $w \psi^{\perp}$, and the cancellation of the curvature $\partial_\theta S$ that turns the bread into the covariance form displayed in (4.6) is carried out, in the equivalent population forms of Remark~\ref{rmk:scaling} in Appendix~\ref{app:proof-optimal-q}.} This proves Proposition~1.

\emph{Insensitivity to $\hat \eta$ (Corollary~1).} Since the displayed identity holds for every $\eta$, the map $\eta \mapsto E_{\mathcal D}\{ I W\, \psi^{\perp}( \theta_0, \eta)\}$ is identically zero, so $G := \partial_\eta E_{\mathcal D}\{ I W\, \psi^{\perp}( \theta_0, \eta)\}\big|_{\eta = \eta^*} = 0$. Expanding instead about $( \theta_0, \eta^*)$ with $\hat \eta$ estimated,
\[
0 = \mathbb P_n \{ w \psi^{\perp}\}( \theta_0, \eta^*) + A_{\tilde \pi}\, ( \hat \theta_n - \theta_0) + G\, ( \hat \eta - \eta^*) + R_n ,
\]
the term $G\, ( \hat \eta - \eta^*)$ vanishes; with the $\sqrt n$-consistency of $\hat \eta$ and $R_n = o_p( n^{-1/2})$, $\sqrt n\, ( \hat \theta_n - \theta_0) \rightsquigarrow N( 0, \Sigma_{\tilde \pi})$, the same limit as with $\eta = \eta^*$ held fixed. The variability of $\hat \eta$ does not enter the $\sqrt n$-scaled limit.

\emph{Step~2: Conditional CLT for the LLB posterior.} Define the empirical measure $\mathbb F_n$ assigning mass $1/n$ to each sampled unit and the LLB-randomised measure $\mathbb F_n^{( j)}$ assigning Dirichlet weight $g_{ji}/n$ to unit $i$. The LLB posterior draw $\theta^{( j)}$ solves $\int w\, \psi^{\perp}( \theta; \hat \eta) \, d \mathbb F_n^{( j)} = 0$, i.e., it is the M-estimator on a Dirichlet-resampled version of the data with per-unit estimating function $w \psi^{\perp}$. By \citet{lo1987}, Theorem~2.1, the Bayesian bootstrap empirical process $\sqrt n \, ( \mathbb F_n^{( j)} - \mathbb F_n)$ converges, conditionally on the data, to a tight Gaussian process with covariance equal to that of the empirical-process limit $\sqrt n \, ( \mathbb F_n - \mathcal F_{\mathrm{samp}})$ under the sample distribution $\mathcal F_{\mathrm{samp}}$.

The map $\mathbb F \mapsto \theta( \mathbb F)$ defined implicitly by $\int w\, \psi^{\perp}( \theta; \hat \eta) \, d \mathbb F = 0$ is Hadamard-differentiable at $\mathbb F_n$ tangentially to the set of bounded mean-zero functions, with derivative $h \mapsto A_{\tilde \pi}^{ -1} h\{ w\, \psi^{\perp}( \theta_0, \hat \eta)\}$ \citep{Vaart:98}, Theorem~3.6{; both $A_{\tilde \pi}$ and $V_{\tilde \pi}$ of (4.6) are moments under $\mathcal F_{\mathrm{samp}}$ itself, so the covariance of the limiting process at the per-unit function $w \psi^{\perp}$ is $V_{\tilde \pi}$ and no scale conversion is needed}. The functional delta method gives
\[
\sqrt n \, ( \theta^{( j)} - \hat \theta_n) \mid \text{data} \;\rightsquigarrow\; N\bigl( 0, \, \Sigma_{\tilde \pi} \bigr) ,
\]
which is the conclusion (4.10). This is the form of the LLB conditional CLT proved by \citet{lyddon2019}, Theorem~1, applied to our specific per-unit M-estimating function $w \psi^{\perp}$.

\emph{Step~3: Profile validity.} The profile construction with $\hat \eta$ held fixed across Dirichlet draws differs from the joint construction (in which $\hat \eta$ would be re-estimated within each draw) only by a second-order term, by the orthogonality property (4.8). Specifically, if $\hat \eta^{( j)}$ denotes the weight-model MLE on the $j$th Dirichlet-resampled data, then $\hat \eta^{( j)} - \hat \eta = O_p( n^{-1/2})$ conditionally on the data, and the resulting perturbation of $\theta^{( j)}$ is $O_p( n^{-1})$ by orthogonality. Both profile and joint constructions therefore have the same limiting conditional distribution. \qed

\subsection{Proof of Proposition~2}
\label{app:proof-optimal-q}

The following remark records the population forms of the sandwich moments used below.

\begin{remark}[Sample-level moments and population forms]
\label{rmk:scaling}
The bread and meat (4.6) are moments under the sample law and are estimated by their direct empirical counterparts, $n^{-1} \sum_{i \in A} w_i q_i\, \{ \cdot\}$ and $n^{-1} \sum_{i \in A} w_i^2 q_i^2\, \{ \cdot\}$ with $\{ \cdot\} = ( S_i - m_{q, i})( S_i - m_{q, i})^\top$; no knowledge of $N$ or of the sampling fraction enters (4.7). By the Horvitz--Thompson identity, the population forms are
\[
A_q = \varrho^{-1}\, E_p\bigl\{ q\, ( S - m_q)( S - m_q)^\top\bigr\}, \qquad
V_q = \varrho^{-1}\, E_p\bigl\{ q^2\, E_p( W \mid x, y)\, ( S - m_q)( S - m_q)^\top\bigr\} ,
\]
with $\varrho = P( I = 1)$; the scalar $\varrho$ multiplies every $\Sigma_q$ identically and so drops out of the comparison of working factors in Proposition~2, which may be conducted in either set of moments. A check under Bernoulli sampling with $\pi \equiv \varrho$ and $q \equiv 1$: for the mean of a $N( \mu, \sigma^2)$ population, $A_q = 1/( \varrho \sigma^2)$ and $V_q = 1/( \varrho^2 \sigma^2)$, so $\Sigma_q = \sigma^2$, the classical iid variance of the sample mean at the $\sqrt n$ rate.
\end{remark}

The sandwich variance $\Sigma_q = A_q^{-1} V_q A_q^{-1}$ and the design-unbiasedness underlying it are given by Proposition~1; it remains to compute the bread and meat in closed form and optimize. Write $r = S( \theta_0) - m_q( x; \theta_0)$, $c( x, y) = E_p( W \mid x, y)$ and $\bar q( x) = E_p\{ q \mid x\}$, so that $E_p\{ q S \mid x\} = \bar q\, m_q$. {By the Horvitz--Thompson identity, the sample-law moments (4.6) equal $\varrho^{-1}$, $\varrho = P( I = 1)$, times the population forms computed below; the common scalar multiplies every $\Sigma_q$ identically and is immaterial to the Loewner comparison, so we compute and compare the population forms, writing $A_q$ and $V_q$ for them within this proof.}

\emph{Bread and meat.} The design $I \mid W \sim \mathrm{Bernoulli}( 1/W)$, $w = W$, gives $E_{\mathcal D}\{ I W \mid x, y\} = 1$ and $E_{\mathcal D}\{ ( I W)^2 \mid x, y\} = E_p( W \mid x, y)$. Since $q\, r$ is a function of $( x, y)$, the meat is immediate, $V_q = E_p\{ q^2 c\, r r^\top\}$. For the bread, $A_q = - \partial_\theta G_q( \theta_0)$ with $G_q( \theta) = E_p\{ q( S( \theta) - m_q( x; \theta))\}$; as $q$ does not depend on $\theta$ and $E_{\mathcal D}\{ I W \mid x, y\} = 1$,
\[
A_q = - E_p\{ q\, \partial_\theta S\} + E_p\{ q\, \partial_\theta m_q\} .
\]
Write $N( x; \theta) = E_p\{ q S \mid x\}$ and $\bar q( x; \theta) = E_p\{ q \mid x\}$, both integrated against $f( y \mid x; \theta)$, so that $m_q = N / \bar q$. Differentiating under the integral with $\partial_\theta f = S f$ gives $\partial_\theta N = E_p\{ q( \partial_\theta S + S S^\top) \mid x\}$ and $\partial_\theta \bar q = E_p\{ q S \mid x\} = N = \bar q\, m_q$, so by the quotient rule
\[
\partial_\theta m_q = \frac{\partial_\theta N}{\bar q} - \frac{N N^\top}{\bar q^{\,2}} = \frac{E_p\{ q( \partial_\theta S + S S^\top) \mid x\}}{\bar q} - m_q m_q^\top .
\]
Because $\partial_\theta m_q$ depends only on $x$, $E_p\{ q\, \partial_\theta m_q\} = E_p\{ \bar q\, \partial_\theta m_q\} = E_p\{ q\, \partial_\theta S\} + E_p\{ q S S^\top\} - E_p\{ \bar q\, m_q m_q^\top\}$, whose first term cancels $- E_p\{ q\, \partial_\theta S\}$. Hence
\[
A_q = E_p\{ q S S^\top\} - E_p\{ \bar q\, m_q m_q^\top\} = E_p\{ q\, ( S - m_q)( S - m_q)^\top\} = E_p\{ q\, r r^\top\} ,
\]
the second equality because $E_p\{ q S \mid x\} = \bar q\, m_q$; this is the population form of (4.6). The curvature $\partial_\theta S$ enters twice with opposite signs and cancels for any model, leaving the $q$-weighted covariance of the centered score.

\emph{Optimization.} Holding the centering fixed, apply the matrix Cauchy--Schwarz inequality to $U = c^{-1/2} r$ and $Z = c^{1/2} q\, r$: since $E_p\{ U U^\top\} = E_p\{ c^{-1} r r^\top\}$, $E_p\{ Z Z^\top\} = V_q$ and $E_p\{ U Z^\top\} = A_q$,
\[
A_q V_q^{-1} A_q \preceq E_p\{ c^{-1} r r^\top\} ,
\]
with equality if and only if $q \propto c^{-1} = \bar \pi$; among working factors the efficient instrument down-weights each unit by its design variance $E_p( W \mid x, y)$, the generalized-least-squares solution. {At $q = \bar \pi$ the population-form bread and meat coincide at $E_p\{ \bar \pi\, r r^\top\} = \varrho\, \mathcal I^\star$, using $c\, \bar \pi^2 = \bar \pi$; equivalently the sample-law moments (4.6) coincide at $\mathcal I^\star = E_s\{ W \bar \pi\, r r^\top\}$, so $\Sigma_{\bar \pi} = ( \mathcal I^\star)^{-1}$.} Finally $\psi_{\bar \pi}$ is the Corollary~C.2 efficient score of \citet{morikawa2025}; every $\psi_q$ is a regular estimating function for the model, so none attains asymptotic variance below the semiparametric bound $( \mathcal I^\star)^{-1}$ realized at $q^\star = \bar \pi$. {Corollary~C.2 there is the $N$-unknown case, matching our use of sampled units only, and \citet{morikawa2025} state their central limit theorems at the $\sqrt N$ scaling over population units with bound $\{ E( S_{\mathrm{eff}}^{\otimes 2})\}^{-1}$; since $E( S_{\mathrm{eff}}^{\otimes 2}) = E_p\{ \bar \pi\, r r^\top\} = \varrho\, \mathcal I^\star$, multiplying the resulting variance by $n/N \to \varrho$ converts it to the $\sqrt n$ scaling used here and returns exactly $( \mathcal I^\star)^{-1}$: the two normalizations agree.} This gives the Loewner-global optimality and absorbs the dependence of $m_q$ on $q$ left open by the Cauchy--Schwarz step. \qed

\subsection{Proof of Proposition~3}
\label{app:proof-crossfit}

Throughout, $\psi^{\perp}( \alpha, q) = q\, \{ S_\alpha( \alpha) - m_q( x; \alpha)\}$ with $m_q( x; \alpha) = E\{ q\, S_\alpha \mid X = x; \alpha\} / E\{ q \mid X = x; \alpha\}$, the joint-score analog of (4.4)--(4.2). Under condition (i) the weights are bounded, $1 \le W \le \bar w$ say, so the sample and population laws of $( X, Y)$ are mutually absolutely continuous with bounded density ratio, and $L_2$ convergence under one implies $L_2$ convergence under the other; and, as in Step~1 of Appendix~\ref{app:proof-thm-procI-bvm}, the sampled units are independent draws from the sample distribution. We write $k( i)$ for the fold containing unit $i$.

\emph{Step 1: exact conditional centering (claim (a)).} Fix $k$ and condition on the training data $\{ ( x_i, y_i, w_i): i \in A \setminus I_k\}$, so that $\hat q_{-k}$ is a fixed positive function taking values in $[ c, C]$. The centering $m_{\hat q_{-k}}$ is by construction the $\hat q_{-k}$-weighted conditional mean of $S_\alpha$ under the analytic model, so the identity (4.5) holds with $S_\alpha$ in place of $S$ and $\hat q_{-k}$ in place of $q$:
\[
E\bigl\{ \hat q_{-k}( X, Y) \bigl[ S_\alpha( \alpha_0; X, Y) - m_{\hat q_{-k}}( X; \alpha_0)\bigr] \bigm| X = x; \alpha_0\bigr\} = 0 \qquad \text{for every } x .
\]
Because the fold assignment is independent of the data and the sampled units are independent, the units in $I_k$ are independent of the training data; applying the Horvitz--Thompson identity exactly as in Appendix~\ref{app:proof-thm-llb} converts the design expectation for a fold-$k$ unit into the displayed analytic-model expectation, giving (5.5). The identity is definitional in the factor---it holds pathwise for every fixed positive function, not merely in a limit---so no $o_p(n^{-1/4})$-type first-order nuisance-rate condition is needed for this exact centering identity itself. Cross-fitting enters only to make $\hat q_{-k}$ fixed relative to the fold on which it is evaluated.

\emph{Step 2: stabilization of the influence function (claim (b)).} Write $\Delta_k = \psi^{\perp}( \alpha_0, \hat q_{-k}) - \psi^{\perp}( \alpha_0, q^*)$ and decompose
\[
\Delta_k = ( \hat q_{-k} - q^*) \{ S_\alpha( \alpha_0) - m_{q^*}\} - \hat q_{-k}\, \{ m_{\hat q_{-k}} - m_{q^*}\} .
\]
Writing $m_q = N_q / \bar q$ with $N_q( x) = E\{ q\, S_\alpha \mid x\}$ and $\bar q( x) = E\{ q \mid x\} \in [ c, C]$, the elementary bound
\[
\| m_{\hat q_{-k}} - m_{q^*}\|( x) \;\le\; c^{-1} E\bigl\{ | \hat q_{-k} - q^*|\, \| S_\alpha( \alpha_0)\| \bigm| x\bigr\} + c^{-2} C\, E\bigl\{ \| S_\alpha( \alpha_0)\| \bigm| x\bigr\}\, E\bigl\{ | \hat q_{-k} - q^*| \bigm| x\bigr\}
\]
follows from the quotient decomposition $m_{\hat q} - m_{q^*} = ( N_{\hat q} - N_{q^*}) / \bar{\hat q} + N_{q^*} ( 1/\bar{\hat q} - 1/\bar q^*)$. By the conditional Cauchy--Schwarz inequality and condition (iii), $E_p\{ \| \Delta_k\|^2 \mid \hat q_{-k}\} \le \kappa_0\, \| \hat q_{-k} - q^*\|_{L_2( P)}^2$ for a constant $\kappa_0$ depending only on $c$, $C$, and $\sup_x E\{ \| S_\alpha( \alpha_0)\|^2 \mid X = x\}$. Now consider the fold-$k$ fluctuation
\[
\mathbb G_k = n^{-1/2} \sum_{i \in I_k} w_i\, \Delta_k( x_i, y_i) .
\]
Conditionally on the training data, $\mathbb G_k$ is a normalized sum of independent terms, each with mean zero by Step~1 (both $\psi^{\perp}( \alpha_0, \hat q_{-k})$ and $\psi^{\perp}( \alpha_0, q^*)$ are exactly centered), and its conditional variance is bounded by a constant multiple of $( | I_k| / n)\, \bar w^2\, \| \hat q_{-k} - q^*\|_{L_2( P)}^2 \to_p 0$ by condition (ii), using the bounded density ratio between the sample and population laws. Chebyshev's inequality conditionally on the training data, followed by dominated convergence, gives $\mathbb G_k = o_p( 1)$; summing over the $K$ (fixed) folds,
\[
n^{-1/2} \sum_{i \in A} w_i\, \psi^{\perp}\bigl( \alpha_0, \hat q_{-k( i)}; x_i, y_i\bigr) = n^{-1/2} \sum_{i \in A} w_i\, \psi^{\perp}( \alpha_0, q^*; x_i, y_i) + o_p( 1) \;\rightsquigarrow\; N( 0, V_{q^*}) ,
\]
the limit being the CLT of Proposition~1 at the fixed factor $q^*$ applied to the joint score. The same envelope bounds give uniform convergence of the Jacobian $n^{-1} \sum_{i \in A} w_i\, \partial_\alpha \psi^{\perp}( \alpha, \hat q_{-k( i)})$ to its population counterpart $-A_{q^*}( \alpha)$ over $\mathcal N$, so the standard M-estimation expansion \citep[][Theorem~5.21]{Vaart:98} yields $\sqrt n\, ( \hat \alpha - \alpha_0) \rightsquigarrow N( 0, A_{q^*}^{-1} V_{q^*} A_{q^*}^{-1})$, which is claim (b).

\emph{Step 3: conditional CLT for the one-step draws (claim (c)).} Conditionally on the data, the per-unit scores $\widehat \varphi_{\alpha i}$ and derivatives $\widehat A_i$ entering (5.4) are fixed. By \citet{lo1987}, Theorem~2.1, the Dirichlet-weighted sum $n^{-1/2} \sum_{i \in A} ( g_{ji} - 1)\, \widehat \varphi_{\alpha i}$ converges conditionally, for almost every data sequence, to a centered Gaussian with covariance $\lim_n n^{-1} \sum_{i \in A} \widehat \varphi_{\alpha i} \widehat \varphi_{\alpha i}^\top$, and Step~2 together with the consistency of $\hat \alpha$ identifies this limit as $V_{q^*}$; likewise $n^{-1} \sum_{i \in A} g_{ji} \widehat A_i \to_p A_{q^*}$ conditionally. Slutsky's theorem applied to (5.4) gives claim (c); the asymptotic equivalence of the one-step draw and the full Dirichlet-weighted solve is the one-step argument already invoked for Algorithm~1 \citep[][Section~5.7]{Vaart:98}.

\emph{Step 4: efficiency at $q^* = \bar \pi$ (claim (d)).} The bread--meat computation of Appendix~\ref{app:proof-optimal-q} used only the score identity $\partial_\alpha f = S_\alpha f$ and the ratio definition of the centering, so it applies verbatim to the joint score: at $q^* = \bar \pi$, $A_{\bar \pi} = V_{\bar \pi} = \mathcal I^\star$ and $\Sigma_{\bar \pi} = ( \mathcal I^\star)^{-1}$, the optimum of the family by Proposition~2. \qed

\subsection{Large-sample theory for Procedure~I}
\label{app:proof-thm-procI-bvm}

Procedure~I is a coherent parametric Bayesian posterior for the sample-level model $f_s$, and its large-sample behavior is governed by standard parametric asymptotics once the sample-level independence induced by the design is made explicit. Write $\vartheta = ( \theta^\top, \eta^\top)^\top$ for the full parameter and $\vartheta_0 = ( \theta_0, \eta_0)$ for its true value.

\begin{theorem}[Posterior consistency and asymptotic efficiency of Procedure~I]
\label{thm:procI-bvm}
Suppose
\begin{enumerate}[label=(\roman*)]
\item the design is Poisson sampling with first-order inclusion probabilities bounded away from $0$ and $1$, with $n/N$ converging to a positive limit;
\item the sample-level joint density $f_s( y, w \mid x; \vartheta)$ of (3.5)--(3.6) is correctly specified, identifiable in $\vartheta$, and differentiable in quadratic mean at $\vartheta_0$;
\item the averaged Fisher information $\bar I_s( \vartheta_0) = E\{ \dot \ell( \vartheta_0) \dot \ell( \vartheta_0)^\top\}$ of the per-unit log-density $\ell( \vartheta) = \log f_s( Y, W \mid X; \vartheta)$ is finite and nonsingular, the expectation taken under the sample distribution;
\item the prior $p_0( \vartheta)$ has a density that is continuous and positive in a neighborhood of $\vartheta_0$.
\end{enumerate}
Then the joint posterior (3.7) satisfies the Bernstein--von Mises property: for almost every realization of the data,
\[
\sup_{B} \Bigl| P\bigl( \sqrt n\, ( \vartheta - \hat \vartheta_n) \in B \mid \mathrm{data}\bigr) - N\bigl( 0, \bar I_s( \vartheta_0)^{-1}\bigr)( B) \Bigr| \;\to_p\; 0 ,
\]
where $\hat \vartheta_n$ is the maximum sample-likelihood estimator. In particular the marginal posterior of $\theta$ concentrates at $\theta_0$ and is asymptotically $N( \hat \theta_n, n^{-1} \Omega_\theta)$ with
\[
\Omega_\theta = \bigl[ \bar I_s( \vartheta_0)^{-1}\bigr]_{\theta\theta} = \bigl( \bar I_{\theta\theta} - \bar I_{\theta\eta} \bar I_{\eta\eta}^{-1} \bar I_{\eta\theta}\bigr)^{-1} ,
\]
the inverse efficient information for $\theta$ in the sample-level model with $\eta$ profiled out. Procedure~I thus attains the model-based efficiency bound for $\theta$ under correct specification of the sample-level model.
\end{theorem}

\emph{Step~1: Sample-level independence.} Under Poisson sampling the indicators $I_i$ are independent across $i$, with $P( I_i = 1 \mid x_i, y_i, w_i) = \pi_i = w_i^{-1}$. Combining the superpopulation density $f( y \mid x; \theta_0)$, the population weight density $f_p( w \mid x, y; \eta_0)$, and Bayes' rule under $I = 1$ as in (3.6), the units retained in the sample are, conditionally on their realized covariates $\{ x_i\}_{i \in A}$ treated as ancillary, independent with density $f_s( y_i, w_i \mid x_i; \vartheta_0)$ \citep{Pfeffermann1998p}. Since $n / N$ converges to a positive limit, $n \to \infty$ almost surely. The Procedure~I likelihood $\prod_{i \in A} f_s( y_i, w_i \mid x_i; \vartheta)$ is therefore a correctly specified likelihood for independent observations.

\emph{Step~2: Bernstein--von Mises for the joint posterior.} Condition (ii) supplies differentiability in quadratic mean of $\vartheta \mapsto f_s( \cdot \mid x; \vartheta)$ at $\vartheta_0$ with score $\dot \ell( \vartheta_0)$, and identifiability of $\vartheta_0$ supplies a uniformly consistent sequence of tests of $\vartheta_0$ against the complements of its neighborhoods. The averaged information of condition (iii) arises as the almost-sure limit $n^{-1} \sum_{i \in A} I_s( \vartheta_0 \mid x_i) \to \bar I_s( \vartheta_0)$ by the law of large numbers over the sample covariate distribution, and is finite and nonsingular. With the prior condition (iv), the parametric Bernstein--von Mises theorem \citep[][Theorem~10.1]{Vaart:98} applies to the independent-observation likelihood of Step~1 and gives
\[
\sup_{B} \Bigl| P\bigl( \sqrt n\, ( \vartheta - \hat \vartheta_n) \in B \mid \mathrm{data}\bigr) - N\bigl( 0, \bar I_s( \vartheta_0)^{-1}\bigr)( B) \Bigr| \to_p 0 ,
\]
where $\hat \vartheta_n$ is asymptotically efficient, $\sqrt n\, ( \hat \vartheta_n - \vartheta_0) = \bar I_s( \vartheta_0)^{-1} n^{-1/2} \sum_{i \in A} \dot \ell( \vartheta_0; y_i, w_i, x_i) + o_p( 1)$.

\emph{Step~3: Marginalization and the efficiency bound.} The marginal posterior of $\theta$ is the $\theta$-marginal of the limiting Gaussian, hence asymptotically $N( \hat \theta_n, n^{-1} [ \bar I_s( \vartheta_0)^{-1}]_{\theta\theta})$. The block-inverse formula gives $[ \bar I_s^{-1}]_{\theta\theta} = ( \bar I_{\theta\theta} - \bar I_{\theta\eta} \bar I_{\eta\eta}^{-1} \bar I_{\eta\theta})^{-1} = \Omega_\theta$, the inverse of the efficient information for $\theta$ in the joint sample-level model with $\eta$ profiled out. This is the Cram\'er--Rao bound for $\theta$ within the parametric sample-level model, so Procedure~I attains the model-based efficiency bound. \qed

The benchmark $\Omega_\theta$ is the \emph{model-based} (parametric) efficiency bound for the assumed sample-level model, and should not be conflated with the semiparametric bound of Section~4. The plug-in posterior discussed in Section~3 shares this limit only when the analytic--weight cross-information $\bar I_{\theta\eta} = E_s\{ \partial_\eta m_{\tilde \pi}( X; \theta_0, \eta_0)\}$ vanishes; in general it conditions on $\hat \eta$, omits the $\theta$--$\eta$ coupling, and is first-order narrower than the joint marginal rather than equivalent to it.

\begin{remark}[Relation to the semiparametric bound]
\label{rmk:procI-vs-procII-eff}
Under correct specification of the weight model, Procedure~I attains the model-based bound $\Omega_\theta$, the Cram\'er--Rao bound within the parametric sample-level model, while Procedure~II carries the sandwich variance $\Sigma_{\tilde \pi} = A_{\tilde \pi}^{-1} V_{\tilde \pi} A_{\tilde \pi}^{-1}$ of its orthogonalized estimating function (Proposition~1), both at the per-sampled-unit $\sqrt n$ scaling. Since the parametric maximum-likelihood estimator is efficient within the assumed model, $\Omega_\theta$ is no larger than the variance of any other regular estimator in that model, so Procedure~I is at least as efficient as Procedure~II under correct specification; the price is that Procedure~I's optimality is contingent on the weight model, whereas Procedure~II remains consistent using only the analytic model. Procedure~II's default variance is not shown here to coincide with the semiparametric bound, although the efficient variant that replaces $\tilde \pi$ by $\bar \pi$ does attain it (Proposition~2). The empirical counterpart is Scenario~A of Section~6, where Procedure~I attains approximately nominal coverage at roughly half the root mean squared error of the design-based methods.
\end{remark}

\section{Implementation details and calculations for the numerical studies}
\label{app:LFS-calcs}

{This appendix collects the linear-Gaussian implementation details for Algorithm~2, used in Sections~6 and~7, and the calculations behind the numerical results of Section~7.}

\begingroup
\subsection{Joint score and centering for the linear-Gaussian model}
\label{app:gaussian-impl}

For the linear Gaussian analytic model, let
\[
\alpha=(\theta_0,\theta_1,\tau)^\top, \qquad \tau=\log\sigma^2, \qquad b(x)=(1,x)^\top,
\]
and define the joint analytic score
\[
S_\alpha(\alpha;x,y)=
\begin{pmatrix}
b(x)\{y-b(x)^\top\theta\}/\sigma^2\\[2mm]
-1/2+\{y-b(x)^\top\theta\}^2/(2\sigma^2)
\end{pmatrix},
\qquad \sigma^2=\exp(\tau).
\]
For a fixed cross-fitted $\widehat{\bar\pi}$, the centering in (5.3) is
\begin{equation}
\widehat m_\alpha(x;\alpha)=
\frac{\int \widehat{\bar\pi}(x,y)S_\alpha(\alpha;x,y)f(y\mid x;\alpha)\,dy}
{\int \widehat{\bar\pi}(x,y)f(y\mid x;\alpha)\,dy}.
\label{eq:joint-mbarpi}
\end{equation}
The integrals in (\ref{eq:joint-mbarpi}) are one-dimensional in the Gaussian model and are evaluated by Gauss--Hermite quadrature.
\endgroup

\subsection{Data and the Gamma GLM weight model}
\label{app:LFS-wmodel}

The data file \texttt{data.txt} contains $n = 142$ rows and four columns: unit identifier, design weight $w_i$, employment count $e_i$, and total payroll (in thousands of dollars) $p_i$. Throughout we write $x_i = \log e_i$ and $y_i = \log( 1000 \, p_i)$, so that $y_i$ is the natural logarithm of payroll in dollars and $x_i$ is the natural logarithm of employment. The sample-level weight model is
\[
W_i \mid x_i, y_i \sim \mathrm{Gamma}\bigl( \mu_i, \nu\bigr) , \qquad \log \mu_i = \beta_0 + \beta_1 x_i + \beta_2 y_i ,
\]
parametrized by the conditional mean $\mu_i = E_s( W \mid x_i, y_i; \beta)$ and a constant dispersion $\nu$. The model is fitted by maximum likelihood using the standard \texttt{glm} routine with \texttt{family = Gamma(link = "log")}. The fitted coefficients are
\[
( \hat \beta_0, \hat \beta_1, \hat \beta_2) = ( 14.478, \, -0.520, \, -0.560) ,
\]
with dispersion $\hat \nu = 1.118$. The implied conditional inclusion probability is $\tilde \pi( x, y; \hat \beta) = \exp\{ -( \hat \beta_0 + \hat \beta_1 x + \hat \beta_2 y)\}$, and the fitted smoothed weights $\hat w_i = \exp( \hat \beta_0 + \hat \beta_1 x_i + \hat \beta_2 y_i)$ range from $9.8$ to about $10\,547$ across the sample and have Pearson correlation $0.66$ with the raw weights $w_i$.

The Gamma GLM with log link directly estimates the conditional mean $E_s( W \mid x, y)$ that the Sverchkov--Pfeffermann identity (3.2) requires, in contrast with ordinary least squares of $\log w_i$ on $( x_i, y_i)$, which estimates the conditional mean of $\log W$. The two estimators differ by an additive bias of order $\frac{1}{2} \mathrm{Var}( \log W \mid x, y)$ if the residuals are approximately log-normal; for the Canadian Workforce data this bias is non-negligible because the raw weights span more than two orders of magnitude. The Gamma GLM is also numerically stable across the observed weight range, where the beta-prime sample-level model (3.10) has a multi-modal likelihood. The framework of Section~3 accommodates any parametric model for $E_s( W \mid x, y)$; we use the Gamma GLM here as a robust default for real-data applications.

\subsection{Procedure~I: exact factorization of the joint posterior}
\label{app:LFS-procI}

For the linear-normal analytic model combined with the log-linear conditional mean $\tilde \pi( x, y; \beta) \propto \exp( -\beta_2 y)$ (treating the $\beta_0 + \beta_1 x$ part as a constant in $y$ given $x$), the conditional log-likelihood at a given value of the weight-model parameter $\beta$ is
\[
\ell_C( \theta, \sigma^2; \beta)
 = \sum_{i \in A} \Bigl\{ \log \tilde \pi( x_i, y_i; \beta)
     + \log \phi\bigl( y_i; \theta_0 + \theta_1 x_i, \sigma^2\bigr)
     - \log Z_i( \theta, \sigma^2; \beta) \Bigr\} ,
\]
where $\phi( y; \mu, \sigma^2)$ is the normal density and the normalizing integral $Z_i = \int \tilde \pi( x_i, y; \beta) \phi( y; \theta_0 + \theta_1 x_i, \sigma^2) \, dy$ has the closed form
\[
\log Z_i( \theta, \sigma^2; \beta) = -\beta_0 - \beta_1 x_i - \beta_2 ( \theta_0 + \theta_1 x_i) + \tfrac{1}{2} \beta_2^2 \sigma^2 ,
\]
obtained by completing the square in the exponential combining the Gaussian density of $y \mid x$ with the log-linear $\tilde \pi$. No quadrature is required for the Canadian Workforce application; the conditional log-likelihood is closed-form in $( \theta_0, \theta_1, \sigma^2)$ for every $\beta$, so the closed form is available at each draw of the joint sampler below.

The closed form of $Z_i$ implies that $Y \mid X = x, I = 1$ is normal with mean $\theta_0 + \theta_1 x - \beta_2 \sigma^2$ and variance $\sigma^2$. Writing $\alpha_0 = \theta_0 - \beta_2 \sigma^2$ therefore factorizes the joint posterior over $( \theta, \beta, \nu)$ into a Gaussian-regression block for $( \alpha_0, \theta_1, \sigma^2)$ and a Gamma-weight-model block for $( \beta, \nu)$. We draw the first block exactly, update the second by adaptive random-walk Metropolis, and transform back to $\theta_0 = \alpha_0 + \beta_2 \sigma^2$ for every draw. Priors are flat on $\alpha_0$, $\theta_1$, and $\beta$, and flat on $\log \sigma^2$ and $\log \nu$, equivalently $p( \sigma^2) \propto 1/\sigma^2$ and $p( \nu) \propto 1/\nu$ on the natural scales. The analysis uses four chains with $5{,}000$ burn-in and $5{,}000$ retained draws per chain. The resulting posterior summaries are $\hat\theta_0^{(\mathrm I)}=9.839$ (SD $0.131$), $\hat\theta_1^{(\mathrm I)}=0.907$ (SD $0.032$), a $95\%$ credible interval $(0.845,0.969)$ for $\theta_1$, and $\hat\sigma^{2,(\mathrm I)}=0.325$, as reported in the Canadian Workforce results table in Section~7 of the main paper.

\subsection{Procedure~II via the orthogonalized LLB}
\label{app:LFS-procII}

For the linear-normal model with parameter vector $\theta = ( \theta_0, \theta_1)$, the analytic-model score is $S( \theta; x, y) = ( y - \theta_0 - \theta_1 x) ( 1, x)^\top / \sigma^2$. The conditional-mean correction of (4.2) is computed under the analytic-model density $f( y \mid x; \theta) = \phi( y; \theta_0 + \theta_1 x, \sigma^2)$, weighted by $\tilde \pi( x, y; \hat \beta) \propto \exp( -\hat \beta_2 y)$. By the same completing-the-square argument as in Appendix~\ref{app:LFS-procI}, the conditional distribution of $Y$ given $X = x$ weighted by $\tilde \pi$ is $N( \theta_0 + \theta_1 x - \hat \beta_2 \sigma^2, \sigma^2)$. The conditional mean of $( Y - \theta_0 - \theta_1 x)$ under this weighted distribution is therefore $-\hat \beta_2 \sigma^2$, giving the closed form
\[
m_{\tilde \pi}( x; \theta, \hat \beta) = -\hat \beta_2 \, ( 1, x)^\top
\]
for the correction (4.2) in this linear-Gaussian / log-linear setting. The correction does not depend on $\theta$, so no linearization point is needed.

The orthogonalized score per sampled unit, with the $1/\sigma^2$ factor absorbed into the proportionality, is
\[
\psi_i^{\perp}( \theta; \hat \beta)
 = \tilde \pi_i \, ( 1, x_i)^\top \, \bigl\{ y_i + \hat \beta_2 \sigma^2 - \theta_0 - \theta_1 x_i \bigr\} ,
\]
where $\tilde \pi_i = \tilde \pi( x_i, y_i; \hat \beta)$. The LLB estimating equation
$\sum_{i \in A} g_{ji} \, w_i \, \psi_i^{\perp}( \theta; \hat \beta) = 0$
is a $2 \times 2$ linear system in $\theta$ with closed-form solution
\[
\theta^{( j)} = ( X^\top D_j X)^{-1} X^\top D_j \tilde y ,
\quad
\tilde y_i = y_i + \hat \beta_2 \hat \sigma^{2,( \mathrm{DB})} ,
\quad
D_j = \mathrm{diag}( g_{ji} \, w_i \, \tilde \pi_i) ,
\]
with $X$ the $n \times 2$ matrix of $( 1, x_i)$ rows and $\hat \sigma^{2,( \mathrm{DB})} = 0.299$ the design-based variance estimate substituted for $\sigma^2$ in the shift. Over $B = 5\,000$ Dirichlet draws, the posterior summaries are
\[
( \hat \theta_0^{( \mathrm{II})}, \hat \theta_1^{( \mathrm{II})}) = ( 9.710, \, 0.948)
\quad \text{with posterior SDs} \quad ( 0.166, \, 0.040) ,
\]
and $\hat \sigma^{2, ( \mathrm{II})} = 0.258$ from the design- and Dirichlet-weighted mean squared error of the fitted residuals.

\subsection{Prior specifications and sampler settings for Procedure~I in the simulation study}
\label{app:procI-sampler}

For Procedure~I in the simulation study of Section~6 we sample the full joint posterior (3.7) of the main paper. The analytic-model coordinates $(\theta_0,\theta_1,\log\sigma^2)$ have flat priors, while the weight-model coordinates use independent weakly informative priors $\beta_j\sim N(0,10^2)$ and $\log\phi\sim N(0,10^2)$. We use an adaptive random-walk Metropolis sampler in ridge-aligned coordinates, combining local updates of the analytic- and weight-model blocks with a seven-dimensional joint proposal to accommodate posterior dependence between the blocks. The integral $Z_i(\theta,\eta)$ is one-dimensional in $y$ given $x_i$ and is evaluated by 21-point Gauss--Hermite quadrature. Each Monte Carlo replicate uses $1{,}000$ burn-in iterations and $1{,}000$ retained draws without thinning. Representative four-chain runs with $2{,}000$ burn-in and $2{,}000$ retained draws per chain gave split-$\hat R<1.05$ for every parameter in both scenarios.


\clearpage
\begin{thebibliography}{32}

\bibitem[Beaumont and Bocci(2009)]{beaumont2009}
Beaumont, J.~F. and Bocci, C. (2009).
A practical bootstrap method for testing hypothesis from survey data.
\emph{Survey Methodology}, 35:25--35.

\bibitem[Binder(1983)]{Binder1983}
Binder, D.~A. (1983).
On the variances of asymptotically normal estimators from complex surveys.
\emph{International Statistical Review}, 51:279--292.

\bibitem[Bissiri et~al.(2016)Bissiri, Holmes, and Walker]{bissiri2016}
Bissiri, P.~G., Holmes, C.~C., and Walker, S.~G. (2016).
A general framework for updating belief distributions.
\emph{Journal of the Royal Statistical Society, Series B}, 78(5):1103--1130.

\bibitem[Chambers and Skinner(2003)]{Chambers2003}
Chambers, R. and Skinner, C.~J.~E. (2003).
\emph{Analysis of Survey Data}. John Wiley \& Sons.

\bibitem[Chernozhukov et~al.(2018)Chernozhukov, Chetverikov, Demirer, Duflo, Hansen, Newey, and Robins]{chernozhukov2018double}
Chernozhukov, V., Chetverikov, D., Demirer, M., Duflo, E., Hansen, C., Newey, W., and Robins, J. (2018).
Double/debiased machine learning for treatment and structural parameters.
\emph{The Econometrics Journal}, 21(1):C1--C68.

\bibitem[Ferrari and Cribari-Neto(2004)]{ferrari2004}
Ferrari, S. and Cribari-Neto, F. (2004).
Beta regression for modelling rates and proportions.
\emph{Journal of Applied Statistics}, 31:407--419.

\bibitem[Fuller(2009)]{fuller2009}
Fuller, W.~A. (2009).
\emph{Sampling Statistics}. Wiley.

\bibitem[Isaki and Fuller(1982)]{isaki1982}
Isaki, C.~T. and Fuller, W.~A. (1982).
Survey design under the regression superpopulation model.
\emph{Journal of the American Statistical Association}, 77:89--96.

\bibitem[Jacob et~al.(2017)Jacob, Murray, Holmes, and Robert]{jacob2017}
Jacob, P.~E., Murray, L.~M., Holmes, C.~C., and Robert, C.~P. (2017).
Better together? Statistical learning in models made of modules.
\emph{arXiv preprint arXiv:1708.08719}.

\bibitem[Kim and Shao(2013)]{kim2013b}
Kim, J.~K. and Shao, J. (2013).
\emph{Statistical Methods for Handling Incomplete Data}. Chapman \& Hall/CRC.

\bibitem[Kim and Skinner(2013)]{kim2013}
Kim, J.~K. and Skinner, C.~J. (2013).
Weighting in survey analysis under informative sampling.
\emph{Biometrika}, 100:358--398.

\bibitem[Kim and Wang(2023)]{kimwang2023}
Kim, J.~K. and Wang, H. (2023).
A note on weight smoothing in survey sampling.
\emph{Survey Methodology}, 49(1):31--38.

\bibitem[Le\'on-Novelo and Savitsky(2019)]{novelo2019}
Le\'on-Novelo, L.~G. and Savitsky, T.~D. (2019).
Fully Bayesian estimation under informative sampling.
\emph{Electronic Journal of Statistics}, 13:1608--1645.

\bibitem[Lo(1987)]{lo1987}
Lo, A.~Y. (1987).
A large sample study of the Bayesian bootstrap.
\emph{Annals of Statistics}, 15(1):360--375.

\bibitem[Lyddon et~al.(2019)Lyddon, Holmes, and Walker]{lyddon2019}
Lyddon, S.~P., Holmes, C.~C., and Walker, S.~G. (2019).
General Bayesian updating and the loss-likelihood bootstrap.
\emph{Biometrika}, 106:465--478.

\bibitem[Monahan and Boos(1992)]{monahan1992}
Monahan, J.~F. and Boos, D.~D. (1992).
Proper likelihoods for Bayesian analysis.
\emph{Biometrika}, 79:271--278.

\bibitem[Morikawa et~al.(2025)Morikawa, Terada, and Kim]{morikawa2025}
Morikawa, K., Terada, Y., and Kim, J.~K. (2025).
Semiparametric adaptive estimation under informative sampling.
\emph{Annals of Statistics}, 53(3):1347--1369.

\bibitem[Neyman(1959)]{neyman1959}
Neyman, J. (1959).
Optimal asymptotic tests of composite statistical hypotheses.
In Grenander, U., editor, \emph{Probability and Statistics: The Harald Cram\'er Volume}, pages 213--234.
Almqvist \& Wiksell, Stockholm.

\bibitem[Pfeffermann et~al.(1998)Pfeffermann, Krieger, and Rinott]{Pfeffermann1998p}
Pfeffermann, D., Krieger, A., and Rinott, Y. (1998).
Parametric distributions of complex survey data under informative probability sampling.
\emph{Statistica Sinica}, 8:1087--1114.

\bibitem[Pfeffermann and Sverchkov(1999)]{Pfeffermann1999}
Pfeffermann, D. and Sverchkov, M. (1999).
Parametric and semiparametric estimation of regression models fitted to survey data.
\emph{Sankhy\=a, Series B}, 61:166--186.

\bibitem[Plummer(2015)]{plummer2015}
Plummer, M. (2015).
Cuts in Bayesian graphical models.
\emph{Statistics and Computing}, 25(1):37--43.

\bibitem[Rao and Molina(2015)]{rao2015}
Rao, J.~N.~K. and Molina, I. (2015).
\emph{Small Area Estimation}. Wiley, 2 edition.

\bibitem[Rubin-Bleuer and Schiopu-Kratina(2005)]{rubin-bleuer2005}
Rubin-Bleuer, S. and Schiopu-Kratina, I. (2005).
On the two-phase framework for joint model and design-based inference.
\emph{The Annals of Statistics}, 33(6):2789--2810.

\bibitem[Savitsky and Toth(2016)]{savitsky2016}
Savitsky, T.~D. and Toth, D. (2016).
Bayesian estimation under informative sampling.
\emph{Electronic journal of Statistics}, 10:1677--1708.

\bibitem[Sugden and Smith(1984)]{Sugden1984}
Sugden, R.~A. and Smith, T.~M.~F. (1984).
Ignorable and informative designs in survey sampling inference.
\emph{Biometrika}, 71:495--506.

\bibitem[Sverchkov and Pfeffermann(2004)]{Sverchkov04}
Sverchkov, M. and Pfeffermann, D. (2004).
Prediction of finite population totals based on the sample distribution.
\emph{Survey Methodology}, 30:79--92.

\bibitem[Tao et~al.(2021)Tao, Mercaldo, Haneuse, Maronge, Rathouz, Heagerty, and Schildcrout]{tao2021two}
Tao, R., Mercaldo, N.~D., Haneuse, S., Maronge, J.~M., Rathouz, P.~J., Heagerty, P.~J., and Schildcrout, J.~S. (2021).
Two-wave two-phase outcome-dependent sampling designs, with applications to longitudinal binary data.
\emph{Statistics in medicine}, 40(8):1863--1876.

\bibitem[van der Vaart(1998)]{Vaart:98}
van der Vaart, A. (1998).
\emph{Asymptotic Statistics}. Cambridge University Press, London.

\bibitem[Wang et~al.(2009)Wang, Scharfstein, Tan, and MacKenzie]{wang2009causal}
Wang, W., Scharfstein, D., Tan, Z., and MacKenzie, E.~J. (2009).
Causal inference in outcome-dependent two-phase sampling designs.
\emph{Journal of the Royal Statistical Society Series B: Statistical Methodology}, 71(5):947--969.

\bibitem[Wang et~al.(2018)Wang, Kim, and Yang]{Wang2018}
Wang, Z., Kim, J.~K., and Yang, S. (2018).
Approximate Bayesian inference under informative sampling.
\emph{Biometrika}, 105:91--102.

\bibitem[Williams and Savitsky(2021)]{williams2021}
Williams, M.~R. and Savitsky, T.~D. (2021).
Uncertainty estimation for pseudo-Bayesian inference under complex sampling.
\emph{International Statistical Review}, 89:72--107.

\bibitem[Wood(2011)]{wood2011}
Wood, S.~N. (2011).
Fast stable restricted maximum likelihood and marginal likelihood estimation of semiparametric generalized linear models.
\emph{Journal of the Royal Statistical Society, Series B}, 73(1):3--36.

\end{thebibliography}
\end{document}